\documentclass[prb,amsmath,amssymb,amsbsy,twocolumn,superscriptaddress,showpacs]{revtex4-1}

\usepackage{amsmath,amssymb}
\usepackage{wasysym}
\usepackage{graphics}
\usepackage[pdftex]{graphicx}
\usepackage{epstopdf}
\usepackage{latexsym}
\usepackage{subfigure}
\usepackage{color}
\usepackage{multirow}
\usepackage[]{natbib}
\usepackage[breaklinks,colorlinks = true,linkcolor = red,urlcolor=cyan,citecolor=red]{hyperref}
\usepackage{times}
\usepackage[abs]{overpic}

\newcommand{\be}{\begin{equation}}
\newcommand{\ee}{\end{equation}}
\newcommand{\bw}{\begin{widetext}}
\newcommand{\ew}{\end{widetext}}
\newcommand{\bea}{\begin{eqnarray}}
\newcommand{\eea}{\end{eqnarray}}
\newcommand{\ba}{\begin{array}}
\newcommand{\ea}{\end{array}}

\newcommand{\la}{\langle}
\newcommand{\ra}{\rangle}
\newcommand{\bS}{\boldsymbol{S}}

\begin{document}

\title{Order-by-disorder and magnetic field response in Heisenberg-Kitaev model on a hyperhoneycomb lattice}
\author{SungBin Lee}
\affiliation{Department of Physics, University of Toronto, Toronto, Ontario M5S 1A7, Canada}
\author{Eric Kin-Ho Lee}
\affiliation{Department of Physics, University of Toronto, Toronto, Ontario M5S 1A7, Canada}
\author{Arun Paramekanti}
\affiliation{Department of Physics, University of Toronto, Toronto, Ontario M5S 1A7, Canada}
\affiliation{Canadian Institute for Advanced Research, Toronto, Ontario, M5G 1Z8, Canada}
\author{Yong Baek Kim}
\affiliation{Department of Physics, University of Toronto, Toronto, Ontario M5S 1A7, Canada}
\affiliation{Canadian Institute for Advanced Research, Toronto, Ontario, M5G 1Z8, Canada}
\affiliation{School of Physics, Korea Institute for Advanced Study, Seoul 130-722, Korea}

\date{\today}
\begin{abstract}
  We study the finite temperature phase diagram of the
  Heisenberg-Kitaev model on a three dimensional hyperhoneycomb
  lattice.  Using semiclassical analysis and classical Monte-Carlo
  simulations, we investigate quantum and thermal order-by-disorder,
  as well as the magnetic ordering temperature. We find the parameter regime where
   quantum and thermal fluctuations favor different magnetic orders, 
  which leads to an additional finite temperature phase transition
  within the ordered phase. This transition, however, 
  occurs at a relatively low temperature and the entropic effects may 
  dominate most of the finite temperature region below 
  the ordering temperature. In addition, 
  we explore the magnetization process in the presence of a
  magnetic field and discover spin-flop transitions which are sensitive to
  the applied-field direction.  We discuss implications of our results
  for future experiments on a hyperhoneycomb lattice system such as the
  recently discovered $\beta$-Li$_2$IrO$_3$.
\end{abstract}
\maketitle

\section{introduction}
\label{sec:intro}



Recent investigations on 5d transition metal oxides have suggested
many interesting spin models and possible candidates of spin-liquid
phases in the Mott insulator regime.  \cite{witczak2013correlated,
  PhysRevB.82.064412, PhysRevLett.108.127203, PhysRevLett.110.076402,
  PhysRevB.83.220403, PhysRevB.86.195131, PhysRevLett.108.127204} One
particular direction has been the study of materials containing
edge-shared iridium-oxygen octahedra: the combination of strong
spin-orbit coupling and correlation effects of iridium's 5d electrons
can lead to an unusual spin model in the Mott limit where both
Heisenberg and Kitaev interactions are present.
\cite{PhysRevLett.102.017205,Kitaev20062,PhysRevB.79.024426} In search
of such a unique spin Hamiltonian in real materials, both theoretical
and experimental studies have extensively focused on the
two-dimensional honeycomb systems Na$_2$IrO$_3$ and Li$_2$IrO$_3$.
\cite{PhysRevLett.105.027204, PhysRevB.84.180407, PhysRevB.82.064412,
  PhysRevLett.108.127203, PhysRevLett.110.076402,
  PhysRevLett.109.197201,PhysRevB.88.035107}

A recent experiment by H. Takagi {\emph{et~al.}}\cite{2013_takagi} on
$\beta$-Li$_2$IrO$_3$ strongly suggests the discovery of a
\textit{three-dimensional} lattice system that may realize the HK model in analogy
to the \textit{two-dimensional} honeycomb lattice. This
\textit{hyperhoneycomb} system contains tri-coordinated Ir$^{4+}$ ions
with edge-shared octahedra and the corresponding HK model 
has been recently studied by
Ref.\onlinecite{lee2013heisenberg,kimchi2013three,nasu2013finite}.  In
the former two works, the zero temperature phase diagram of the HK model 
was studied, while in the last work, the strongly anisotropic Kitaev model was
explored.


The HK model on the 3D hyperhoneycomb lattice is distinct from the 2D
case in several ways and hence offers an exciting new platform for the
study of the interplay between spin-orbit and correlation effects.
Firstly, unlike the 2D honeycomb, the magnetic ordering temperature
can be finite for the pure Heisenberg models in three-dimensional
lattices.\cite{PhysRevB.88.024410} Second, when the classical ground
state is degenerate, both thermal and quantum fluctuations may lift
the degeneracy and select a particular ordered state, which is the
well-known order by disorder (ObD) mechanism.  For the case of a 2D
honeycomb lattice, both thermal and quantum fluctuations lift the
accidental degeneracy of the classical ground state and favor the same
ordered state.  They both select spins pointing along cubic
directions, resulting in a six-fold degeneracy of the spin order.
\cite{PhysRevB.88.024410} In general, however, the origin of these two
fluctuations---one from zero-point quantum fluctuation energy and
another from entropy effects---are different and they could favor
different states.
For example, in the $J_1$-$J_2$ Heisenberg model on a diamond lattice,
it has been suggested that thermal and quantum ObD could favor
different states among classically degenerate spiral
orders.\cite{bergman2007order,bernier2008quantum} Another example of
the competition between quantum and thermal ObD is studied in the
antiferromagnetic Heisenberg model on a pyrochlore lattice in the
presence of Dzyaloshinskii-Moriya
interactions.\cite{chern2010pyrochlore}

In this paper, we study the finite temperature phase diagram of the HK
model on the 3D hyperhoneycomb lattice.  Within the classical limit,
there are four distinct magnetic orders; ferromagnet (FM), N\'eel,
skew-stripy, and skew-zig-zag.  We find that both quantum and thermal
fluctuations lift the classical degeneracy in the ground state
manifold and choose particular collinear spin states in every phase.
In a certain parameter regime, quantum and thermal ObD compete and
favor different collinear spin states.  This interplay between quantum
and thermal fluctuations leads to a finite temperature phase
transition within the ordered phase in addition to the paramagnetic
phase transition.
In addition to ObD phenomena, magnetic field effects are also
investigated as a guide for future experiments.  To be specific, we
computed the magnetization process as a function of magnetic field for
each phase and discuss how the Heisenberg and Kitaev spin exchange
interactions can be extracted from the magnetic field response of the
system.

The rest of the paper is organized as follows.  In Sec.\ref{sec:HK},
we start by introducing the HK model on the 3D hyperhoneycomb lattice.
We briefly summarize the zero-temperature phase diagram studied in
Ref.\onlinecite{lee2013heisenberg} and describe key features of HK
model.  In Sec.\ref{sec:finite-T-pd}, we present a finite temperature
phase diagram of the HK model based on both classical Monte Carlo (MC)
simulations and a semi-classical calculation using Holstein-Primakoff
linear spin-wave theory.  These complementary techniques facilitate
the study of a wide range of temperatures and allow us to estimate the
magnetic ordering temperature as well as investigate the competition
of thermal and quantum ObD.  The magnetic field effect is studied in
Sec.\ref{sec:magnetic-field}.  Using MC simulations, we find a
magnetization jump at the saturation field due to a spin-flop
transition for certain regions in the phase diagram.  We also estimate
the saturation field as a function of the ratio of Kitaev interaction
to Heisenberg interaction.  For comparison, we discuss the field
effect within the semi-classical Holstein-Primakoff linear spin-wave
theory.  We present the angle-averaged saturation field values as well
as a more detailed look at the role of applied-field direction.  In
Sec.\ref{sec:discussion}, we summarize our results and discuss
predictions that can be tested against future experiments on a
hyperhoneycomb lattice systems such as $\beta$-Li$_2$IrO$_3$.

\section{Heisenberg-Kitaev model on a hyperhoneycomb lattice}
\label{sec:HK}
The HK model on the hyperhoneycomb lattice in the context of
$\beta$-Li$_2$IrO$_3$ is described in detail in
Ref.\onlinecite{lee2013heisenberg}.  Here we recapitulate several
points relevant to our current work.

The idealized hyperhoneycomb lattice is composed of edge-shared oxygen
octahedra with Ir$^{4+}$ ions at their centers.  Each Ir site is connected
to three other nearest-neighbor Ir ions, thereby generating a
tri-coordinated 3D lattice.  Since each 5d Ir ion is situated in an
octahedral crystal-field environment and atomic spin-orbit coupling is
large, the low-energy physics may be described by $j_{\text{eff}}=1/2$
states.\cite{kim06032009} In the presence of Hund's coupling, the
strong-coupling limit of such a system with edge-shared octahedra can
be described by a HK model with $j_{\text{eff}}=1/2$ pseudospins at
each Ir site.\cite{PhysRevLett.102.017205} The magnetization operator
$\mathbf{M}=\mathbf{L}+2\mathbf{S}$ (taking the Land\`{e} g-factor of
the electron to be 2) projected into the $j_{\text{eff}}=1/2$ subspace
yields $\mathbf{M}=-2\mathbf{J}_{\text{eff}}$ and hence we can treat
the $j_{\text{eff}}=1/2$ pseudospins at each Ir site as spins with an
effective g-factor of
$g_{\text{eff}}=-2$.\cite{witczak2012topological}

The HK model is
\begin{equation}
  {H}_{\rm{HK} } = J \sum_{\la  i j \ra} \bS_i \cdot \bS_j - K \sum_{\la i j \ra, \alpha - \rm{links}}
  S_i^\alpha S_j^\alpha,
\end{equation}
where the first and second terms are the Heisenberg and Kitaev
exchanges respectively.  Here, $\la i j \ra$ indicates
nearest-neighbors $i$ and $j$, and $\alpha$-links $(\alpha= x,y,z)$
denote the three bonds in the tri-coordinated hyperhoneycomb lattice.

The zero-temperature classical phase diagram contains four magnetic
phases.  For the antiferromagnetic Heisenberg region ($J>0$), two
phases are found: the N\'{e}el ($K/|J|<1$) and skew-stripy
($K/|J|>1$).  For the ferromagnetic Heisenberg region ($J<0$), two
other phases are found: the skew-zig-zag ($K/|J|<-1$), and the
ferromagnet (FM, $K/|J|>-1$).  Phase transitions occur between the
N\'{e}el and skew-stripy phases at $J>0, K/|J|=1$, as well as between
the skew-zig-zag and ferromagnet at $J<0, K/|J|=-1$.  As we will see
in Sec. \ref{sec:finite-T-pd}, these phase boundaries are pinned to
these specific $K/|J|$ values and extend to finite temperatures.

Similar to the 2D honeycomb HK model, a four-sublattice rotation maps
$J\rightarrow -J$, $K\rightarrow
K-2J$.\cite{khaliullin2005orbital,PhysRevLett.105.027204,lee2013heisenberg}
By this exact transformation, the N\'{e}el and skew-zig-zag states map
onto each other, as do the skew-stripy and FM states.  Moreover, the
pure antiferromagnetic Heisenberg point maps to the special point
$J<0,K/|J|=-2$ while the pure ferromagnetic Heisenberg point maps to
$J>0,K/|J|=2$, implying that the latter is exactly solvable and that
all four of these points have exact \textit{SU(2)} degeneracy.

\begin{figure}
  \centering
  \setlength\fboxsep{0pt}
  \setlength\fboxrule{0.0pt}
  \fbox{\begin{overpic}[scale=.11]{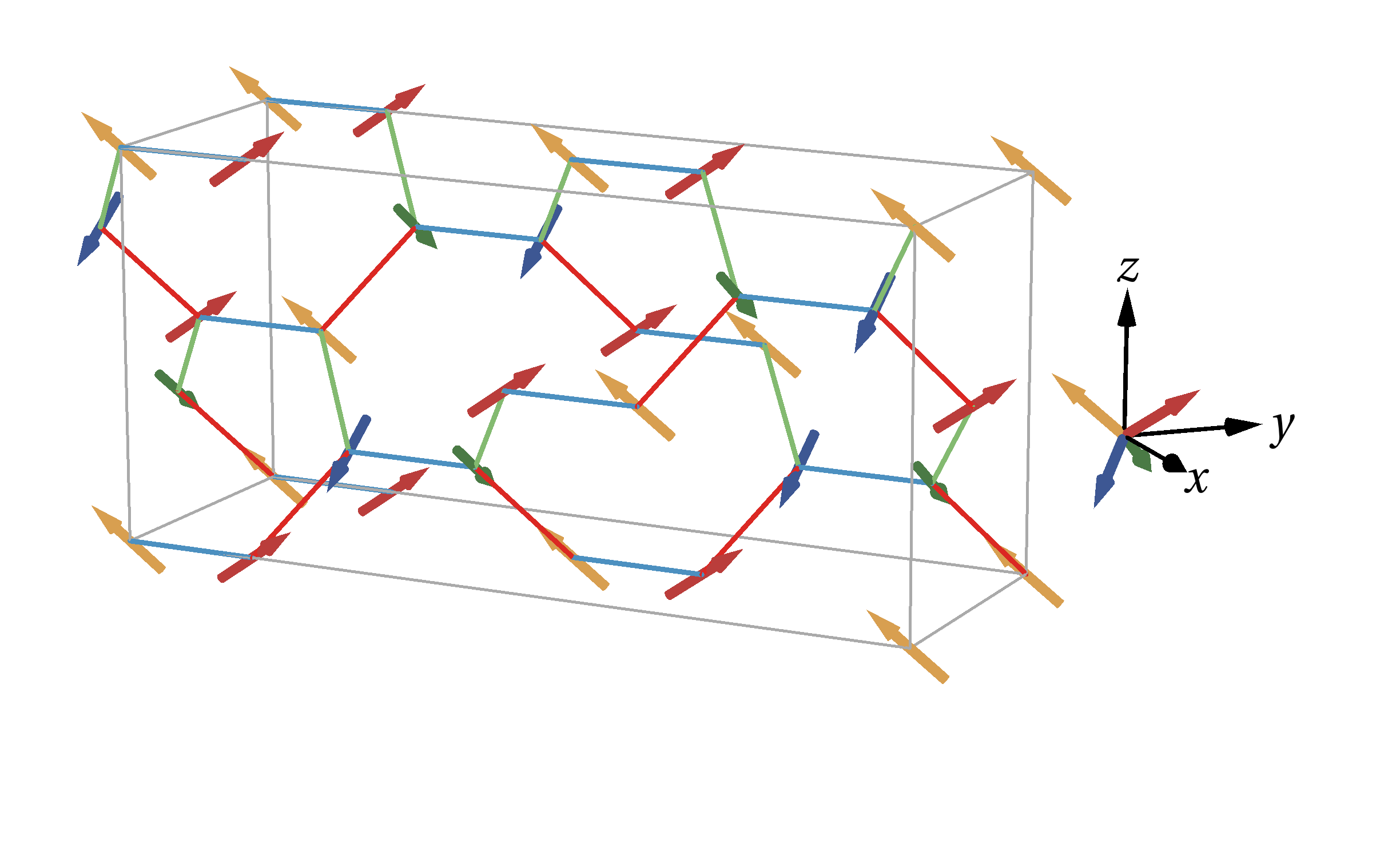}
    \end{overpic}}
  \caption{(color online) The $(111)$-skew-stripy phase on the
    hyperhoneycomb lattice with the coordinate system shown on the
    right.  Each Ir site is described by a localized
    $j_{\text{eff}}=1/2$ pseudospin and is connected with three
    nearest-neighbors.  The $\alpha$-links are colored red, green, and
    blue for the $x$-, $y$-, and $z$- bonds respectively.  The red,
    orange, green, and blue pseudospins point in the $(1,1,1)$,
    $(-1,-1,1)$, $(-1,1,-1)$, and $(1,-1,-1)$ directions.  Note that
    these pseudospins are non-coplanar and form an $8$-sublattice
    structure, which is a generic feature of the skew-stripy and
    skew-zig-zag phases.  On the other hand, the $(100)$, $(010)$, and
    $(001)$ skew-stripy/skew-zig-zag states (not shown) are collinear
    spin orders, while $(n_x n_y 0)$, $(0 n_y n_z)$, and $(n_x 0 n_z)$
    skew-stripy/skew-zig-zag states (not shown) are coplanar states.}
  \label{fig:111stripy}
\end{figure}

Away from these four special points that possess an exact
\textit{SU(2)} symmetry, the HK model possess an \textit{accidental
  SU(2)} symmetry at the zero-temperature classical
level.\cite{PhysRevLett.105.027204,lee2013heisenberg} Since these
accidentally degenerate ground state manifolds will be of primary
concern in our discussion of both order-by-disorder and magnetic field
response, we now detail the parametrization of the degenerate
manifolds used in this work.  In the ferromagnetic and N\'{e}el
regions, the accidental \textit{SU(2)} symmetry can be parametrized by
a 3-vector representing the collinear order.  For example, the
$(100)$-, $(010)$-, $(001)$-, and $(111)$-FM states, also denoted as
$x$-, $y$-, $z$, and $(111)$-FM, have every spin directed in the $x$,
$y$, $z$, and $111$ directions respectively (see
Fig.\ref{fig:111stripy} for definitions of the $x$, $y$, and $z$
axes).  Similarly, we also parametrize a general skew-zig-zag
(skew-stripy) state with a 3-vector $\vec{n}=(n_x,n_y,n_z)$, which
refers to the state obtained by performing a four-sublattices rotation
on the $\vec{n}$ N\'{e}el (FM) state.  In our discussion of ObD, we
will see that the $x$, $y$, and $z$ states play a central role.

Due to the sublattice-dependence of the four-sublattices
transformation, the skew-stripy and skew-zig-zag states are
non-coplanar in general, as the $(111)$-skew-stripy state exemplifies
in Fig.\ref{fig:111stripy}.  In contrast, certain states within the
degenerate manifolds are coplanar or even collinear.  In particular,
the $(100)$-, $(010)$-, and $(001)$-skew-stripy/skew-zig-zag states
are collinear spin orders, while the $(n_x n_y0)$-, $(0 n_y n_z)$-,
and $(n_x 0 n_z)$-skew-stripy/skew-zig-zag states are coplanar states.
As we will see in Sec.\ref{sec:magnetic-field}, this has important
consequences in the magnetic field response of our model.

The low symmetry of the hyperhoneycomb lattice only ensures that
magnetic phases related by a $C^{\hat{x}+\hat{y}}_2$ rotation are
symmetry-related (this can be contrasted with the 2D honeycomb
lattice, where a $C_6$ axis is present).  In particular, the
$z$-N\'{e}el state is distinct from the $x$- and $y$-N\'{e}el states,
but the latter two are symmetry related by the aforementioned
$C^{\hat{x}+\hat{y}}_2$ rotation (the same is true for the
skew-zig-zag, skew-stripy, and FM states).

\section{Finite temperature phase diagram}
\label{sec:finite-T-pd}
Following the zero-temperature phase diagram in
Ref.\onlinecite{lee2013heisenberg}, we explore the finite temperature
phase diagram of HK model.  
As discussed in the previous section, there are four distinct ordered phases : 
FM, N\'eel, skew-stripy, skew-zig-zag. 
Using both MC simulation and a
semi-classical analysis, we provide magnetic ordering temperature and
particular states favored by quantum-, thermal-ObD in the HK model. For
comparison, we show two different phase diagrams: one obtained from MC
simulation in Sec.\ref{subsec:mc} and another obtained from
Holstein-Primakoff bosons in Sec.\ref{subsec:HP-boson}.  We found, in
general, the 3D HK hyperhoneycomb model has higher ordering
temperatures than the 2D honeycomb lattice as expected.  
In addition, it turns out that both quantum and thermal fluctuations always favor 
collinear spin states on every phase. 
However, unlike the 2D honeycomb lattice, there is a certain parameter regime 
where quantum ObD and thermal ObD compete with each other and 
such competition results in an additional  
phase transition below the ordering temperature. 
Nevertheless, this additional phase transition occurs at the very low temperature, 
and for most of the temperature range below ordering temperature, 
the magnetic order chosen by thermal ObD is dominant.

Thermal ObD selects two distinct collinear spin states 
depending on the parameter regime.
When the Heisenberg and Kitaev interactions are frustrated 
(for the different signs of $J$ and $-K$) 
in both unrotated and rotated basis, 
the system prefers collinear spin states with spins pointing along (001) direction. 
Otherwise, it prefers collinear spin state with spins pointing along (100) direction 
or its symmetry equivalent (010) direction.
These two types of magnetic order have different nature of phase transitions.
For the former case, it belongs to the 3D Ising
universality class, allowing collinear spins pointing either $(001)$
or $(00\bar{1})$ direction.  On the other hand, for the latter  
case, there are four symmetry-related magnetic orders with spins
pointing along $(100), (\bar{1}00), (010), (0\bar{1}0)$.  The domain
wall energy for a sharp domain wall 
between $(100)$ order and $(\bar{1}00)$ order is estimated to be 
$D_{x,\bar{x} } = (-3J +K)$ (per bond), which is the same as $D_{y,
  \bar{y}}$ on symmetry grounds.  However, the domain wall energy 
  between $(100)$ order and $(010)$ order 
  is estimated to be $D_{x,y}=
(-3J+K)/2 = D_{x,\bar{x} } /2 $ and is the same as $D_{x,\bar{y} } =
D_{\bar{x},y} = D_{\bar{x},\bar{y}}$. 
Hence, the phase transition of (100)-state (or symmetry equivalent (010)-state) 
is expected to be equivalent to the one of 3D XY model with $Z_4$ anisotropy.
This $Z_4$ anisotropy in 3D XY model is dangerously irrelevant 
and this is studied in Refs.\onlinecite{caselle1998stability,PhysRevB.48.1291,
  PhysRevLett.99.207203,PhysRevE.68.046107}. 
 Thus, the finite temperature transition belongs to the 3D XY universality class. 
 Below, we will discuss the magnetic order and ordering temperature within MC simulation 
 and linear spin wave approximation, but will not attempt to numerically extract 
 the critical exponents at the transition.
  

\subsection{Classical Monte Carlo simulation}
\label{subsec:mc}
\begin{figure}[t]
  \scalebox{0.8}{\includegraphics{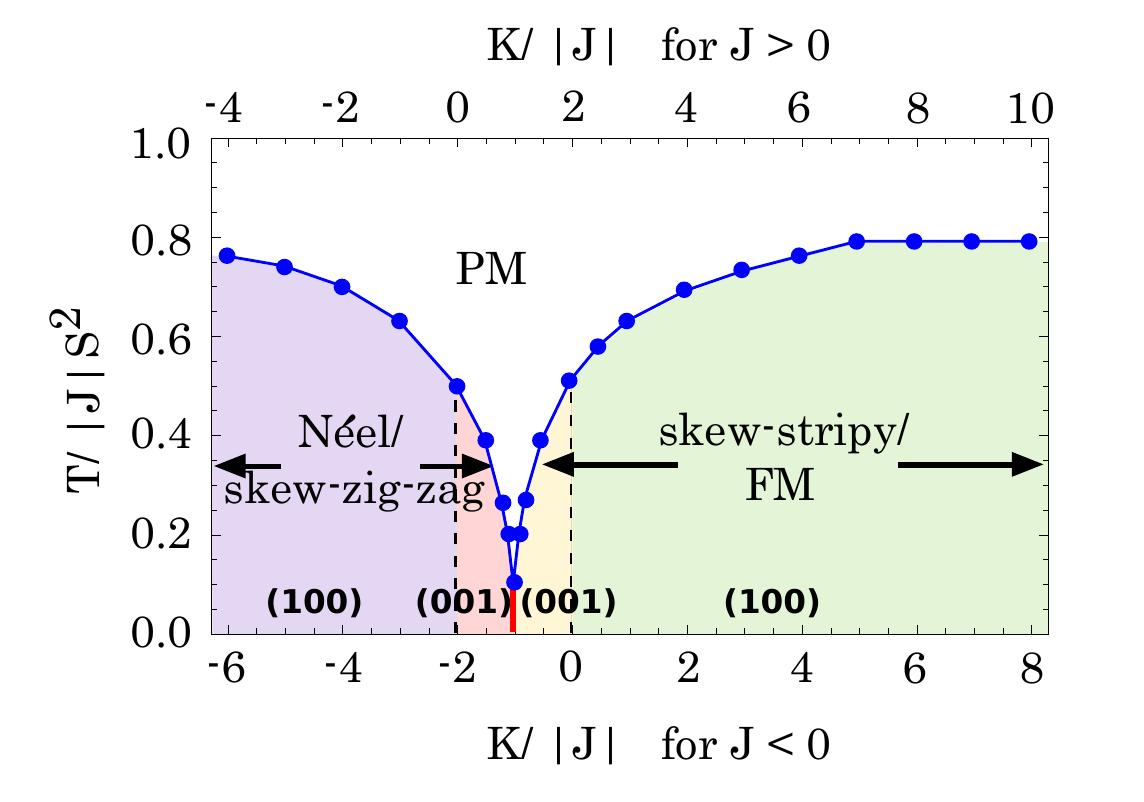}}
  \caption{(color online) Finite temperature phase diagram of classical HK model
    obtained within MC simulation.  For both cases of FM $J$ ($J<0$,
    lower horizontal axis) and AF $J$ ($J>0$, upper horizontal axis),
    there are two magnetic orders : FM and skew zig-zag phase for FM
    $J$, N\'eel and skew-stripy phase for AF $J$.  Blue line
    represents magnetic ordering temperature and red solid line is the
    phase boundary where the two distinct phases are degenerate.  Two
    black dotted lines are for the exact solvable (anti)-ferromagnetic
    Heisenberg model.  Thermal ObD favors collinear FM / skew zig-zag phase (or N\'eel
    / skew-stripy phase) with spins pointing along (001)
    direction, when $J$ and $-K$ have opposite signs (\emph{i.e. frustrated case}) 
    in both unrotated and rotated basis. Otherwise,
    ordered phases have collinear spin states with spins pointing along (100) 
    direction. The errors in the calculated ordering temperatures 
    are smaller than the size of indicated data points.}
  \label{fig:1}
\end{figure}

In this section, we study the finite temperature phase diagram of the
classical HK model using MC simulations based on the standard
Metropolis algorithm.  In our simulations, we treat the spins as three
dimensional unit vectors, $\bS_i = (S_i^x , S_i^y, S_i^z)$ , $|\bS_i
|=1$.  For every data point, we use $10^6$ number of MC sweeps and
thermalization to simulate the system size up to $L=12$ for the
total number of sites $N = 4 L^3$.  Fig.\ref{fig:1} shows the finite
temperature phase diagram as a function of $K/|J|$ for both FM $J$ 
($J<0$) and AF $J$ ($J>0$).  We emphasize that there is one to one mapping 
between FM $J$ and AF $J$, based on a four-sublattice rotation introduced in
Sec.\ref{sec:HK}.  Such basis rotations map the ferromagnetic (FM)
phase to the skew-stripy phase, and the skew-zig-zag phase to the
N\'eel phase (and vice versa).  In Fig.\ref{fig:1}, the upper
horizontal axis is for AF $J$ and the lower horizontal axis is
for FM $J$.  For FM $J$, there are two distinct phases: the FM
phase and the skew zig-zag phase.  The energy crossing between these 
two states occurs at $K/|J| = -1$.  For AF $J$, there are also two
phases: the N\'eel phase and the skew-stripy phase.  The phase
transition occurs at $K/|J| = 1$.  Since a four-sublattice rotation
maps $J \rightarrow -J$ and $K \rightarrow K-2J$, the phase diagram of
FM $J$ can be made consistent with the one for AF $J$ by shifting $K/|J|$ by a
constant of 2.  The blue line indicates the ordering temperature of
the classical HK model.  Unlike the case of the 2D honeycomb lattice,
the ordering temperature is finite even at pure Heisenberg points 
$K/|J| = 0$ for $J>0$ and $J<0$ 
and their equivalent points $K/|J| = \pm2$ for $\pm J >0$ 
after a four-sublattice rotation.  
The phase transition between N\'eel phase and skew-stripy phase (or the
phase transition between skew zig-zag phase and FM phase) occurs at
$K/|J|=1$ (or $K/|J| =-1$) with ordering temperature $T_c/|J| \approx
0.1$.  The ordering temperature is largely suppressed at this
transition point due to the magnetic frustration.  Two black dotted
lines are for the exactly solvable Heisenberg points; pure
(anti-) ferromagnetic Heisenberg model at $K/|J| =0$ 
for $J>0$ and $J<0$
and their equivalent points $K/|J| = \pm2$ for $\pm J >0$ 
after a four-sublattice rotation. 


There are two distinct regimes in every ordered phase (FM, N\'eel,
skew-stripy and skew zig-zag) where thermal ObD selects collinear spin
states with spins pointing along either the (001) or (100) direction.
To confirm such thermal ObD effect in MC simulation, we have looked at
the histogram for each component of uniform (staggered) magnetization
in the FM (N\'eel) phase.  In the (001)-FM phase, for example, the
histogram of $z$ component of uniform magnetization shows a sharp peak
at non-zero values which vary as a function of temperature, whereas,
both $x$ and $y$ components of magnetization both have a peak at
zero. Once we find thermal ObD effect for either FM phase or N\'eel
phase, one-to-one mapping between FM $J$ and AF $J$ automatically
tells us that similar entropic selection for the skew-stripy and skew-zig-zag
phases occur as well.  
We notice that the thermal fluctuations select collinear spin states 
with spins pointing along (001) direction, when the Heisenberg and Kitaev 
interactions are \emph{frustrated} ({\emph i.e.} the signs of $J$ and $-K$ are different)
in both unrotated and rotated basis.
Otherwise, the system favors the collinear spin states with 
spins pointing along (100) direction or its symmetry 
equivalent (010) direction. 
The free energy
calculation from fluctuating fields at the Gaussian level also shows such global selection by
entropy effect.  (See Appendix.\ref{app:free-energy} for details)


%
%

\subsection{Linear spin wave theory}
\label{subsec:HP-boson}

\begin{figure}[hbc]
\centering
\setlength\fboxsep{0pt}
\setlength\fboxrule{0.0pt}
\fbox{\includegraphics[scale=.1,clip=true,trim=0 0 0 0]{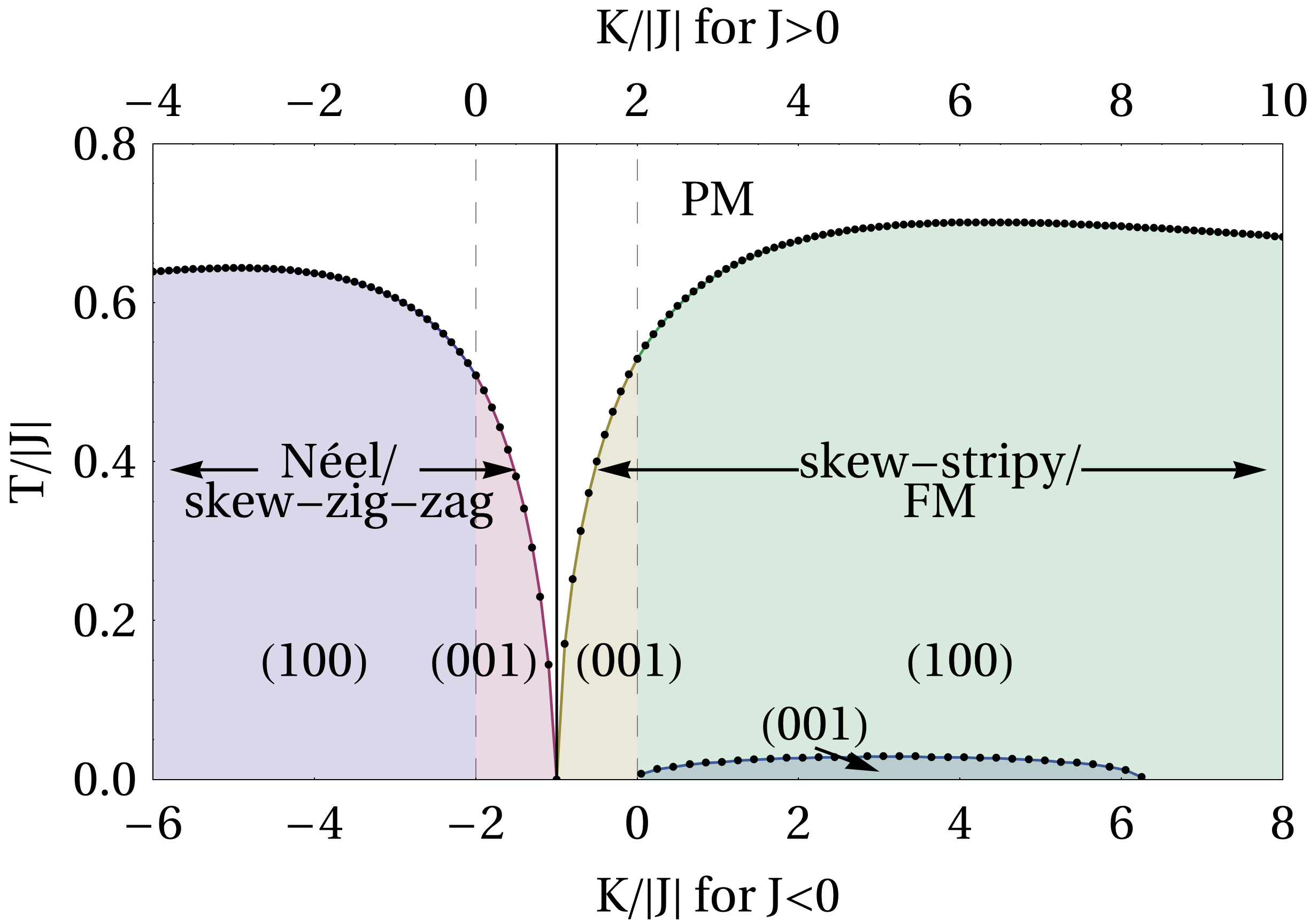}}
\caption{(color online) Finite temperature phase diagram obtained
  within the Holstein-Primakoff linear spin-wave approximation with
  $S=1/2$.  The top and bottom horizontal axes are for the
  antiferromagnetic $(J>0)$ and ferromagnetic $(J<0)$ Heisenberg
  exchanges respectively.  These two axes are related by a
  four-sublattices rotation (see Sec. \ref{sec:HK}).  The N\'{e}el and
  skew-stripy labeling of the left and right regions apply to the top
  axis $(J>0)$ while zig-zag and ferromagnet (FM) apply to the bottom
  axis $(J<0)$.  The first-order phase boundary between these regions
  is indicated by a solid vertical line.  Phase boundaries between
  different states selected by thermal ObD within each region are
  indicated by dashed lines.  The first-order phase boundary between
  the thermal ObD-selected and quantum ObD-selected states appears in
  the lower right of the phase diagram. The estimated ordering
  temperature is given by the curve between the paramagnetic (PM)
  region and the magnetically ordered regions.  See main text for
  details on the ObD mechanism and the nature of the selected states.}
\label{fig:hp_pd}
\end{figure}

\subsubsection*{Competition between thermal and
  quantum order-by-disorder}
Thermal order-by-disorder selects states within the classically
degenerate ground state manifold that have the largest
\textit{thermal} fluctuations: an entropic effect.  In contrast,
quantum order-by-disorder selects states with largest \textit{quantum}
fluctuations and occurs at sufficiently low temperatures when entropic
effects are negligible.  Since the selection criteria are different
between these two mechanisms, the states that are preferred by each
can also differ.  If the selected ground states do differ and are
described by different order parameters, a first-order phase
transition is expected at finite temperature.  To investigate such a
possibility, we analyze the quantum and thermal spin fluctuations at
zero and finite temperature about the classical ground state via
Holstein-Primakoff linear spin-wave theory.  As we will see, the
results of this section will also complement those obtained in our
classical Monte Carlo simulations (see Sec.\ref{subsec:mc}).

Parameterizing the classical degeneracy by $\vec{n}$ (see Sec.\ref{sec:HK}),  
the $\vec{n}$-dependent contribution to the free energy can be written as 
\begin{equation}
  \label{eq:freeEnergyHp}
  \delta F(\vec{n}, T)= T \sum_{\boldsymbol{k}} \log\left( 1 - e^{-\omega_{\boldsymbol{k}}(\vec{n})/T} \right)
    + \frac{1}{2} \sum_{\boldsymbol{k}} \omega_{\boldsymbol{k}}(\vec{n}) ,
\end{equation}
where $T$ is temperature and
$\omega_{\boldsymbol{k}}(\vec{n})$ is the linear spin-wave spectrum
that is dependent on $\vec{n}$ (the band index has been suppressed).
In computing the linear spin-wave dispersion, magnon-magnon
interactions have been ignored and only quadratic terms have been
retained.  We have also taken $S=1/2$ to represent the
$j_{\text{eff}}=1/2$ pseudospins of
$\beta\text{-}\text{Li}_2\text{Ir}\text{O}_3$ in the strong-coupling
limit.

By minimizing the free energy in respect to $\vec{n}$ at both zero and
finite temperatures, we can examine the interplay of quantum ObD and
thermal ObD.  In Eq.\eqref{eq:freeEnergyHp}, the first term is the free
energy of the thermal fluctuations while the last term is the
zero-point quantum fluctuation energy correction to the classical
energy.  Competition between thermal ObD and quantum ObD can occur if
the second and last terms possess different global minima in the
ground state manifold.  A first-order transition will be present if
these minima do not possess symmetry-related order parameters.  For
concreteness, let us now explore this possibility in the N\'{e}el and
skew-stripy region of the phase diagram---the results for the
skew-zig-zag and ferromagnetic regions can be obtained via the
four-sublattices rotation introduced earlier (the phase diagram in
Fig.\ref{fig:hp_pd} shows $K/|J|$ axes for both $J>0$ and $J<0$).

\paragraph*{N\'{e}el $(J>0,K/|J|<1)$:} Both quantum ObD and thermal ObD select the
same state for a given $K/|J|$.  The $x$-N\'{e}el is selected for
$K/|J|<0$ while the $z$-N\'{e}el is selected for $0<K/|J|<1$.  This is
shown in Fig.\ref{fig:hp_pd}.  Both the selected states and the phase
boundary at $K/|J|=0$ match those obtained via our classical Monte
Carlo simulations.  We note that at the pure antiferromagnetic
Heisenberg point, \textit{SU(2)} symmetry is restored and the
accidental degeneracy of the ground state manifold becomes exact hence
no ObD is present.

\paragraph*{Skew-stripy $(J>0,K/|J|>1)$:} When $1<K/|J|<2$, both
quantum ObD and thermal ObD select the $z$-skew-stripy phase, which is
in agreement with our classical Monte Carlo results.  On the other
hand, for $2<K/|J|\lesssim8.3$, quantum ObD selects the
$z$-skew-stripy phase, but thermal fluctuations prefer the
$x$-skew-stripy phase.  At temperatures of the order $T\approx
0.03~J$, a first-order phase transition between these two skew-stripy
phases is present, as seen in Fig.\ref{fig:hp_pd}.  At
$T\gtrsim0.03~J$, the states selected by thermal ObD as well as the
phase boundary at $K/|J|=2$ match the results of our classical Monte
Carlo results.  We again note that the $K/|J|=2$ point is the pure
ferromagnetic Heisenberg point in the rotated basis, hence
\textit{SU(2)} symmetry is restored and no ObD mechanism is present.

\subsubsection*{Ordering temperature}
Using linear spin-wave theory, we can also estimate the magnetic
ordering temperatures and compare the results with those estimated
from our Monte Carlo simulations.  We define the ordering temperature
as the temperature at which the order parameter (local magnetization)
vanishes.  Despite the breakdown of linear spin-wave theory when
fluctuations are large, this definition would serve as a rough
upper-bound of the critical temperature between the ordered phase and
the paramagnetic phase.  Due to the four-sublattices rotation, this
estimate of the ordering temperature applies equally to both
antiferromagnetic ($J>0$) and ferromagnetic ($J<0$) Heisenberg
exchanges.  In Fig.\ref{fig:hp_pd}, the calculated ordering
temperatures as a function of $K/|J|$ is shown.  We remark that
we cannot make a quantitative comparison between the ordering temperature
found within classical Monte Carlo and within this spin-wave
calculation (the former method considers rigid spins of
length $|S|=1$ and incorporates nonlinear interactions 
while the latter describes quantum mechanical spins
with $S=1/2$ within linear spin wave approximation), 
the general trend of decreasing ordering temperature
near the phase boundary between the skew-stripy and N\'{e}el region at
$K/|J|=1$ is consistent between the two calculations (likewise for the
phase boundary between the FM and skew-zig-zag region at $K/|J|=-1$).
Linear spin-wave theory estimates that the ordering temperature
approaches zero as $K/|J|$ approaches this phase boundary.  However,
the computed fluctuations in the local magnetization becomes
comparable to the local magnetization itself, signaling the breakdown
of linear spin-wave theory, hence the estimated ordering temperature
may be significantly modified by magnon interactions at this phase
boundary.

\
\section{Magnetic field response}
\label{sec:magnetic-field}

The magnetic field response will be useful to clarify the nature of
the magnetic order in real materials since the response depends
sensitively on the ordering of the spins.  In this section, we study
the magnetic field response based on both MC simulation in
Sec.\ref{subsec:mc_mag} and a semi-classical analysis in
Sec.\ref{subsec:hp_mag}.  Except for the FM case where the field
effect is trivial, we estimate the saturation field and its
directional dependence in each phase.  For the skew-stripy and
skew-zig-zag phases, the spin-flop transition is present for general
field directions.  This originates from the fact that the general
skew-stripy and skew-zig-zag phases are neither collinear nor
coplanar. For future experiments, we also compute the angle-averaged
saturation field on each ordered phase.

\subsection{Classical Monte Carlo}
\label{subsec:mc_mag}
Simple analysis for the saturation field has already been reported in
Ref.\onlinecite{lee2013heisenberg}.  Here, we extend that study by investigating 
the directional
dependence of the magnetic field response on each phase at finite temperature
using MC simulation.  Including magnetic field $\boldsymbol{h}$, the
Heisenberg-Kitaev model can be written as
\bea {H}_{\boldsymbol{h}} =
J \sum_{\la i j \ra} \bS_i \cdot \bS_j - K \sum_{\la i j \ra, \alpha
  - \rm{links}}
S_i^\alpha S_j^\alpha - \sum_i \boldsymbol{h} \cdot \bS_i.  \nonumber  \\
\label{eq:3}
\eea
Here, the magnetic field $\boldsymbol{h}$ includes the Land\`e
$g$-factor.  Thus, for a given spin magnitude $S$, the magnetic field
is scaled as $g/ S$.  
Unlike our zero-field results, we will need to examine the saturation
field of each magnetic phase separately since the four-subalttices
rotation transforms a uniform field in the unrotated basis to a
site-dependent field in the rotated basis.
We also note that, in principle, 
the (001)- and (100)- states within the same phases 
show different magnetic field response 
when the field strength is comparable to the free energy difference 
between those two states ($\Delta {F}/T \sim 10^{-4}$). 
However, when the field strength is substantially larger than the free energy difference, 
the magnetic field response of these two phases will be indistinguishable.

\paragraph*{N\'eel phase:}
\label{subsubsec:neel-mag}
When the magnetic field is applied in the N\'eel phase,
the uniform magnetization is developed along the field direction,  
having the N\'eel order on its basal plane.
This results in spin canting and  
the energy in terms of canted spin angle $\theta$ out of N\'eel phase is
\be
\label{eq:analyticalHsatNeel}
E_{\boldsymbol{h}}^{\text{N\'eel}} = (\frac{3J}{2} - \frac{K}{2}) (
\sin^2{\theta} -\cos^2{\theta}) - h \sin{\theta},
\ee
The saturation
field is $h_{\rm{sat}} = 6J-2K$ where the canting angle $\theta$
becomes $\pi/2$.  

\begin{figure}[t]
\scalebox{0.5}{\includegraphics{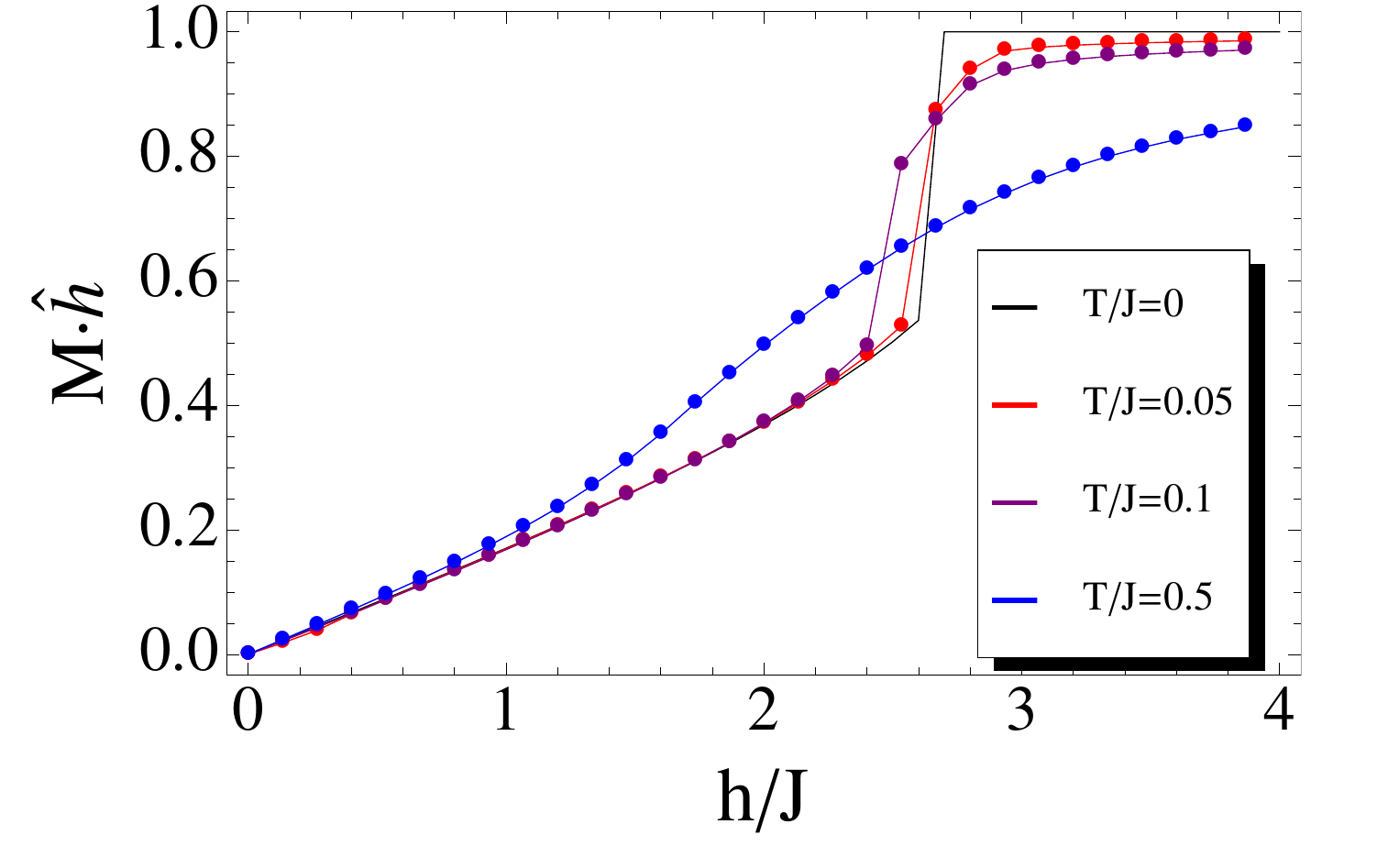}}
\caption{(color online) Magnetization curve as a function of $h/J$ for  $\boldsymbol{h} = h/\sqrt{3} (111)$, 
$K/J = 4$ and $J>0$ (in (100)-skew-stripy phase). Red, purple, blue points are MC results for $T/J =0.05,0.1,0.5$ 
(lines are drawn as a guide to the eye) and 
black line is from numerical energy minimization of $H_{\boldsymbol{h}}$ for $T/J=0$. 
The spin-flop transition is present at the saturation field $h_{\rm{sat}} /J \approx 2.7$. }
\label{fig:2}
\end{figure}

\begin{figure}[t]
\scalebox{0.45}{\includegraphics{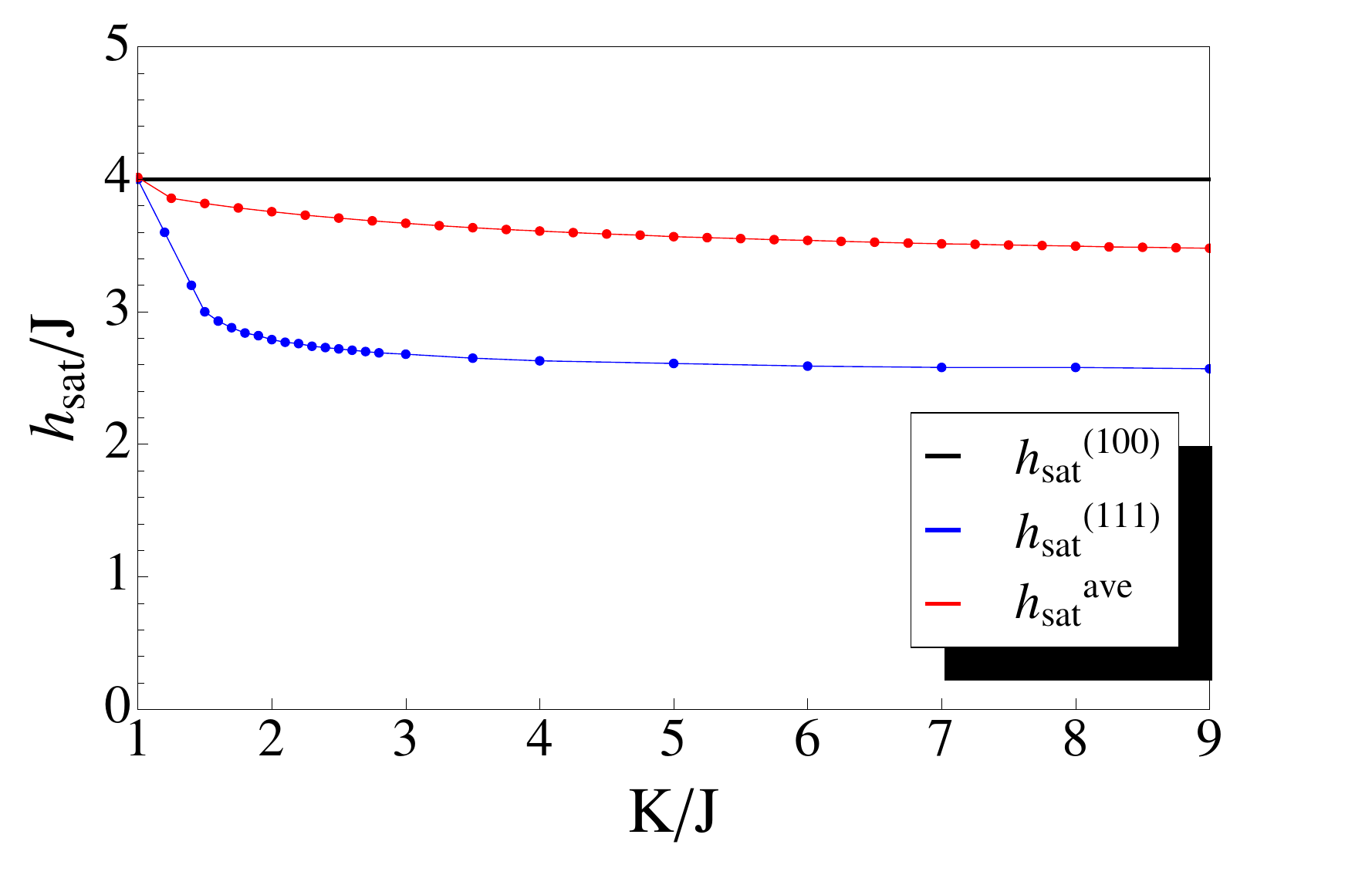}}
\caption{(color online) Plot of saturation field $h_{\rm{sat}}/J$ at zero temperature 
as a function of $K/J$ within the skew-stripy phase. 
Black and blue lines are the saturation field when the field is applied in the
(100), (111) directions respectively, 
based on the energy minimization of $H_{\boldsymbol{h}}$ at indicated points.
Red line is the angle averaged saturation field obtained by the linear spin-wave calculation.
}
\label{fig:3}
\end{figure}

\paragraph*{Skew-stripy phase:} 
\label{subsubsec:stripy-mag}
When the magnetic field is applied perpendicular to the spin
directions of either the $(001)$- or the $(100)$-skew-stripy phases,
the uniform magnetization is developed along the field direction,
resulting in spin canting similar to the N\'eel phase.  The energy in
terms of canted spin angle $\theta$ out of skew-stripy phase is
\be
E_{\boldsymbol{h}= h(100)}^{\rm{stripy}} = \frac{J}{2} ( 3
\sin^2{\theta} -\cos^2{\theta}) -\frac{K}{2} - h \sin{\theta}.  
\ee
In this case, the saturation field is $h_{\rm{sat}}^{(100)} = 4J$ which is
independent of Kitaev interaction $K$.
On the other hand, when the magnetic field direction is not perpendicular 
to the direction of the spin order in the skew-stripy phase, say $\boldsymbol{h} = h /\sqrt{3} (111)$, 
there is no way to develop uniform magnetization along the field direction 
and have skew-stripy phase on its perpendicular plane.
In principle, the skew-stripy phase belongs to neither collinear nor coplanar spin order 
except for special cases like $(001)$-,$(100)$-skew-stripy phases 
where both cases have collinear spin order.
Therefore, the skew-stripy phase cannot be established perpendicular to the field direction 
in general.
For a small field, (001)-,(100)-skew-stripy phase is still robust 
and the system develops uniform magnetization 
perpendicular to its direction. 
In the presence of a large magnetic field, however, the Zeeman energy overcomes 
the spin exchange energy and the skew-stripy phase is eventually destroyed and 
all spins are polarized along the field direction. 
Fig.\ref{fig:2} shows the magnetization as a function of magnitudes $h$ for  
$\boldsymbol{h} = h/\sqrt{3} (111)$ at $K/J =4$ and $J>0$. 
Red, purple and blue points are MC results (lines are drawn as a guide to the eye) 
for finite temperatures $T/J = 0.05,0.1$ and $0.5$ respectively.
Black line is the magnetization curve obtained from energy minimization of 
$ H_{\boldsymbol{h}}$ for zero temperature $T/J=0$.
As we expected, the (100)-skew-stripy phase is still present and uniform magnetization is developed 
perpendicular to the (100) direction for a small field $h<h_{\rm{sat}}$.
At the saturation field $h_{\rm{sat}} /J \approx 2.7$, 
the spins are suddenly all polarized along the field direction, 
resulting in a magnetization jump.
Such a magnetization jump corresponds to the spin-flop transition from (100)-skew-stripy phase  
to fully polarized spin orders along the field direction $\boldsymbol{h} //(111)$. 
Fig.\ref{fig:3} shows the saturation field $h_{\rm{sat}}/J$ for three different cases 
as a function of $K/J$ in the skew-stripy phase. 
Black line is the saturation field when the field is applied along (100) direction.
As discussed before, the saturation field is $h_{\rm{sat}}^{(100)} = 4J$, 
which is independent of the Kitaev interaction $K$.
Blue line is the saturation field $h_{\rm{sat}}^{(111)}$ when the field direction is along (111),
obtained by the energy minimization of $H_{\boldsymbol{h}}$ in Eq.\eqref{eq:3}.
In this case, since the skew-stripy phase is stabilized by the interplay between 
AF Heisenberg interaction ($J>0$) and FM Kitaev interaction ($K>0$), 
$h_{\rm{sat}}^{(111)}/J$ varies near the phase boundary $K/J = 1$ 
but eventually saturates for large $K/J$ to $h_{\rm{sat}}/J \approx 2.6$.
Red line is the angle-averaged saturation field $h_{\rm{sat}}^{\rm{ave}}/J$, 
obtained by the linear spin-wave calculation. (See Section.\ref{subsec:hp_mag})

%
%
%
%

\begin{figure}[t]
\scalebox{0.5}{\includegraphics{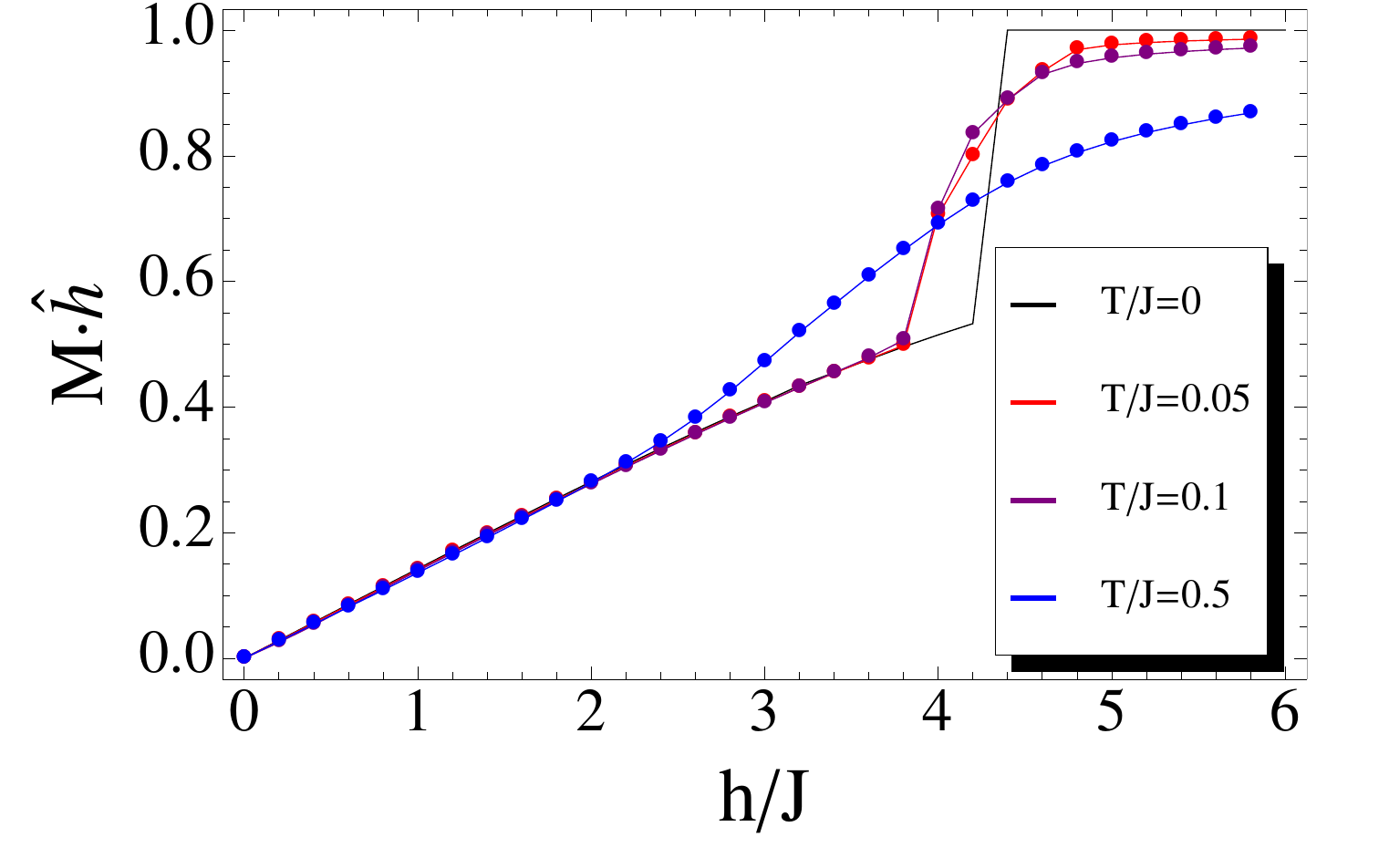}}
\caption{(color online) Magnetization curve as a function of $h/J$ for  $\boldsymbol{h} = h/\sqrt{3} (111)$, 
$K/|J |= -4$ and $J<0$ (in (100)-skew-zig-zag phase). Red, purple, blue points 
are MC results (lines are drawn as a guide to the eye) and  
black line is from numerical energy minimization of $H_{\boldsymbol{h}}$ for $T/J=0$. 
The spin-flop transition is present at the saturation field $h_{\rm{sat}} /J \approx 4.3$. }
\label{fig:4}
\end{figure}

\begin{figure}[t]
\scalebox{0.45}{\includegraphics{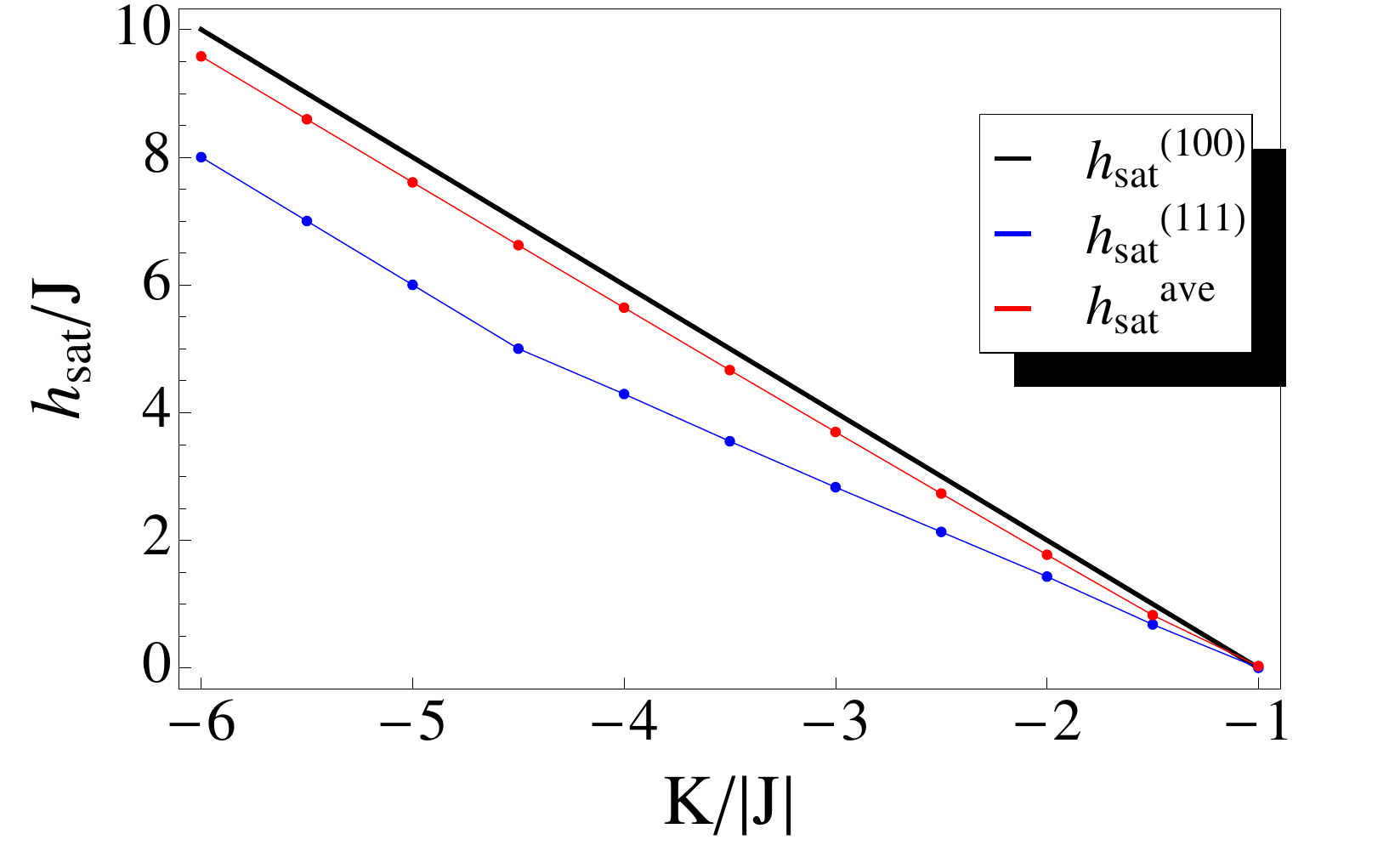}}
\caption{(color online) Plot of saturation field $h_{\rm{sat}}/J$ at zero temperature 
as a function of $K/|J|$ within the skew-zig-zag phase. 
Black and blue lines are the saturation field when the field is applied to 
(100), (111) directions respectively, 
based on the energy minimization of $H_{\boldsymbol{h}}$ at indicated points.
Red line is the angle averaged saturation field obtained by linear spin-wave calculation.}
\label{fig:5}
\end{figure}

\paragraph*{skew-zig-zag phase:} 
\label{subsubsec:zigzag-mag}
In the skew-zig-zag phase, AF Kitaev interaction competes with Zeeman energy 
unlike the case for the skew-stripy phase and 
the saturation field does depend on both Heisenberg and 
Kitaev interactions even when the field is applied perpendicular to the
collinear spin order.
For the field perpendicular to the spin order in (001)-,(100)-zig-zag phase, 
the energy in terms of canted spin angle $\theta$ out of skew-zig-zag phase is
\be
E_{\boldsymbol{h}= h(100)}^{\rm{zig-zag}} = ( \frac{J}{2} + \frac{K}{2}) \cos^2{\theta}
+ ( \frac{3J}{2} - \frac{K}{2}) \sin^2{\theta} - h \sin{\theta}.
\ee
This leads the saturation field to be
$h_{\rm{sat}} = 2J-2K$ for the field perpendicular to the spin order in (001)-,(100)-zig-zag phase.
Similar to the case of the skew-stripy phase, 
the skew-zig-zag phase does not belong to either collinear or coplanar spin order 
in general.
Hence, for general directions of magnetic field, one can expect a similar 
magnetization jump like the case of the skew-stripy phase.
Fig.\ref{fig:4} shows the magnetization as a function of magnetic field 
$h/J$ for $\boldsymbol{h} = h/\sqrt{3}(111)$ at $K/|J| =-4$ and $J<0$. 
The spin-flop transition is present at $h_{\rm{sat}}/J  \approx 4.3$. 
For $h<h_{\rm{sat}}$ , the uniform magnetization is linearly increasing 
with increase of magnetic field $h$. 
We note that the eight sublattices magnetic order is stabilized to minimize 
both AF Kitaev spin interaction and Zeeman energy. 
Fig.\ref{fig:5} represents the saturation field for three different cases 
as a function of $K/|J|$ within the skew-zig-zag phase.
Black line is for the saturation field when the field is along (100) direction, 
 $h_{\rm{sat}}^{(100)} = 2J-2K$.
Blue line is for the saturation field applied in the (111) direction, 
obtained by energy minimization of $H_{\boldsymbol{h}}$ in Eq.\eqref{eq:3}. 
Red line is for the angle-averaged saturation field, $h_{\rm{sat}}^{\rm{ave}}$, 
obtained by linear spin-wave calculation. (See Section.\ref{subsec:hp_mag})
Unlike the skew-stripy phase, the saturation field $h_{\rm{sat}}/J$ increases 
as a function of $|K/J|$.

%
%

%
%

\subsection{Linear spin wave theory}
\label{subsec:hp_mag}

The saturation field at zero temperature can also be computed within
linear spin-wave theory.  By applying a sufficiently large field, the
classical ground state is ferromagnetically ordered in the direction
of the magnetic field.  We then consider the spin-wave spectrum about
such an ordered state and lower the magnetic field strength until the
spin-wave spectrum becomes gapless.  Further decrease in the
applied-field strength will render the classical ferromagnetic state
unstable, indicating that the saturation field has been reached.

Although the magnetization curve and spin-flop transitions revealed in
the previous section could not be obtained within this approach,
linear spin-wave theory allows us to efficiently compute
$h_{\text{sat}}$ with arbitrary field directions and perform
angle-averages, which are of particular experimental interest for
single-crystal, polycrystalline, and powder samples.

Similar to Sec.\ref{subsec:mc_mag}, we set the spin magnitude $S$ and
Land\`{e} g-factor to be arbitrary: the magnetic field is in units of
$g/S$.

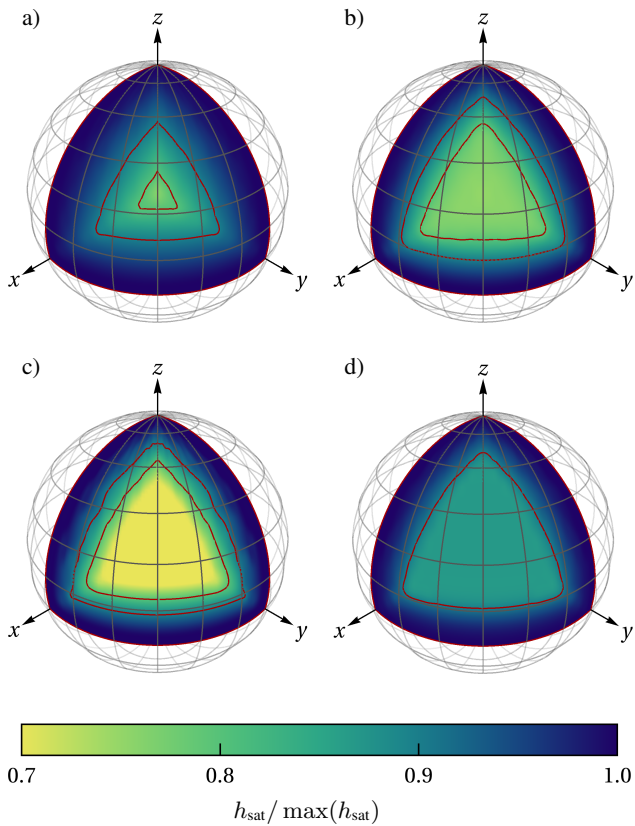
\begin{figure}
  \centering
  \setlength\fboxsep{0pt}
  \setlength\fboxrule{0.0pt}
  \fbox{\begin{overpic}[scale=.1]{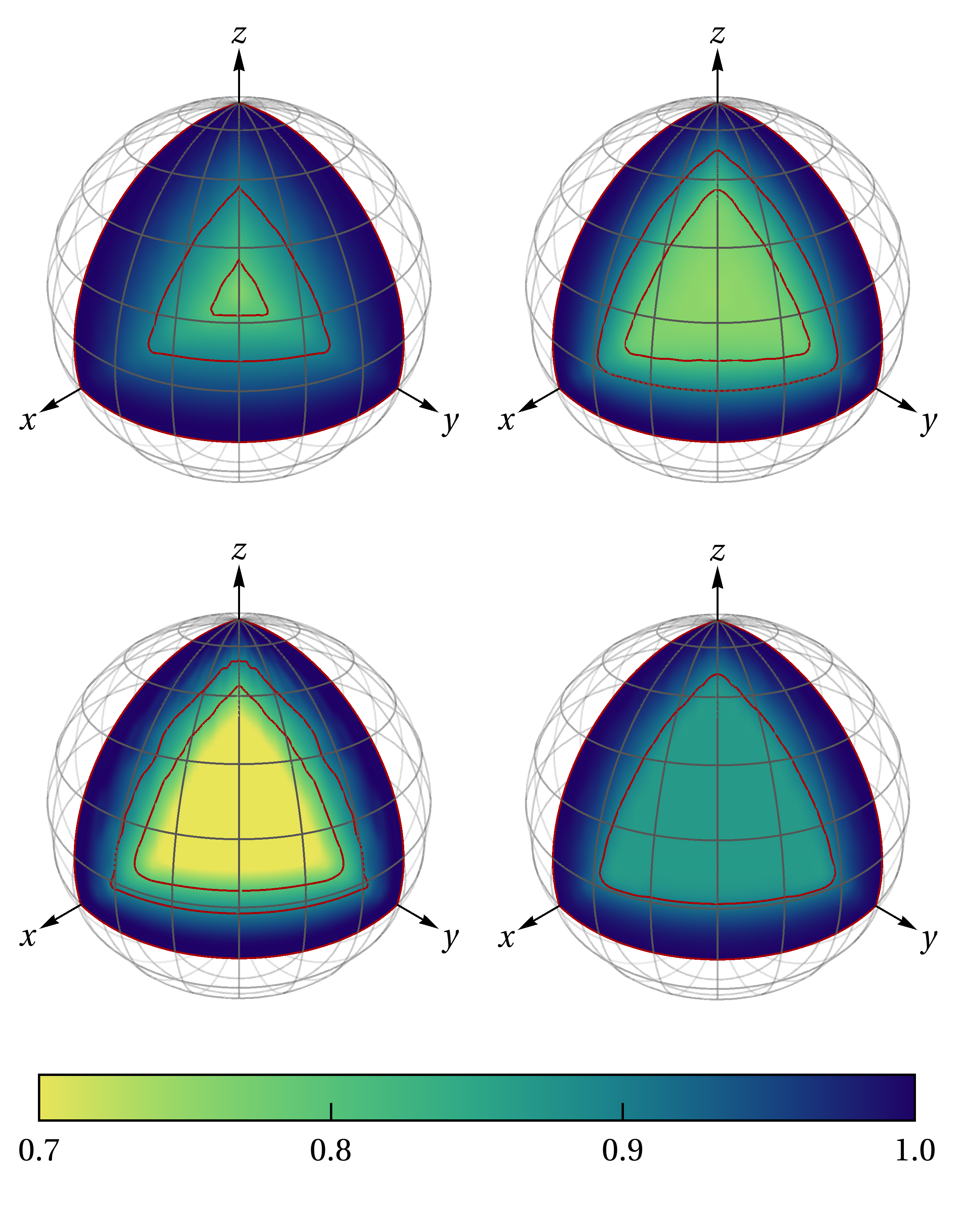}
      \put(10,305){a)}
      \put(132,305){b)}
      \put(10,173){c)}
      \put(132,173){d)}
      \put(90,05){$h_{\text{sat}}/\max(h_{\text{sat}})$}
    \end{overpic}}
  \caption{(color online) Field-direction dependence of saturation
    field ($h_{\text{sat}}$).  The color maps indicate the fraction of
    the saturation field compared to the maximum attained in
    directions perpendicular to the cubic axes.  Red contours are
    shown indicating $h_{\text{sat}}/\max(h_{\text{sat}})$ values of
    0.9 and 0.8.  Figures a) and b) are for the skew-stripy phase at
    $K=1.5|J|$ and $K=3.0|J|$ respectively.  Notice the solid angle
    with low saturation field grows as $|K|$ increases.  Figures c) and
    d) are for the skew-zig-zag phase at $K=-2.0|J|$ and $K=-4.0|J|$.
    Notice the opposite trend: the solid angle with low saturation
    field decreases as $|K|$ increases.}
  \label{fig:hsatK}
\end{figure}

\paragraph*{N\'{e}el ($J>0,K/|J|<1$):} All the N\'{e}el states within
the ground state manifold are collinear, hence no spin-flop transition
is expected at field strengths larger than the quantum zero-point
energy splitting (which is smaller than $10^{-3}|J|$).  This also
implies that the saturation field does not depend on the applied
direction.  The angle-averaged saturation field found is exactly
$h^{\text{ave}}_{\text{sat}}=6J - 2K$, which agrees with the
expression obtained from Eq.\eqref{eq:analyticalHsatNeel}.

\paragraph*{Skew-stripy ($J>0,K/|J|>1$):} For fields applied
perpendicular to the collinear skew-stripy order,
$h_{\text{sat}}=h^{(100)}_{\text{sat}}=4~J$ is \textit{independent} of
$K/|J|$.  In Fig.\ref{fig:hsatK}a) and \ref{fig:hsatK}b), we also see
that the saturation field is sensitive to the field direction due to
the spin-flop transition and it is \textit{reduced} when the field is
applied in the $\langle 111\rangle$-direction with
$h_{\text{sat}}\approx0.75 h^{(100)}_{\text{sat}}=3~J$.  The $K/|J|$
dependence is manifested in the solid angle where the saturation field
is reduced: Fig.\ref{fig:hsatK}a) ($K/|J|=1.5$) has a smaller solid
angle with reduced $h_{\text{sat}}$ compared to Fig.\ref{fig:hsatK}b)
($K/|J|=3.0$).  This implies that the angle-averaged saturation field
will decrease as $K/|J|$ increases, which can be seen in
Fig.\ref{fig:3} ($h^{\text{ave}}_{\text{sat}}$ approaches $3.4~J$
with increasing $K/|J|$).  We also note that the
$h^{\text{ave}}_{\text{sat}}$ is largely \textit{independent} of
$K/|J|$.

\paragraph*{Skew-zig-zag ($J<0,K/|J|<-1$):} The saturation field of
the skew-zig-zag phase also possesses directional dependence due to
the general non-coplanar nature of the skew-zig-zag phase that causes
the spin-flop jump in magnetization.  This is readily seen in Fig.
\ref{fig:hsatK}c and Fig.\ref{fig:hsatK}d.  Contrasting with the
skew-stripy phase, as $|K/J|$ increases, the region with reduced
saturation field decreases in size.  This can be seen by comparing
Fig.\ref{fig:hsatK}c ($K/|J|=-2.0$) with Fig.\ref{fig:hsatK}d
($K/|J|=-4.0$).  The angle-averaged saturation field is shown in
Fig.\ref{fig:5} and is below the value of
$h^{(100)}_{\text{sat}}=2(J - K)$ due to the reduction in
$h_{\text{sat}}$ for general field directions.  As $|K/J|$ increases,
$h^{\text{ave}}_{\text{sat}}$ approaches $h^{(100)}_{\text{sat}}$.

\begingroup
\squeezetable
\begin{table}[htp]
  \begin{ruledtabular}
    \begin{tabular}{c c c}
      Magnetic order & Parameter region &  $\Theta_{\rm{CW}}/ h^{\text{ave}}_{\text{sat}}$ \\
      \hline
      N\'{e}el & $J>0,K/|J|<1$ & $-1/2$ \\
      Skew-stripy & $J>0,K/|J|>1$ & $ (K/J-3)/4$ \\
      Skew-zig-zag & $J<0,K/|J|<-1$& $ (K-3J)/(2J-2K)$
    \end{tabular}
  \end{ruledtabular}
  \caption{ Table of the ratio of Curie-Weiss temperature $\Theta_{\rm{CW}}$ 
    to angle-averaged saturation field $h^{\text{ave}}_{\text{sat}}$ as 
    estimated from linear spin-wave theory.  The saturation field $h^{\text{ave}}_{\text{sat}}$ is scaled by $g/S=4$.  
    For the skew-stripy and skew-zig-zag phases, the quoted result of  
    $\Theta_{\rm{CW}}/h_{\rm{sat}}^{\rm{ave}}$ is for fields along (100), however, this is a 
    good approximation to the angle-averaged result (which is slightly smaller in magnitude). }
  \label{tab:1}
\end{table}
\endgroup

\section{Discussion and Summary}
\label{sec:discussion}

Related to future experiments on the material, we suggest estimating
the magnitudes of the Heisenberg and Kitaev interactions $J,K$ based
on our theory.  From experiments, one can measure the Curie-Weiss
temperature $\Theta_{\rm{CW}}$, ordering temperature $T_c$, and
saturation field $ h_{\rm{sat}} $.  In the high temperature limit, the
Curie-Weiss temperature is given by $\Theta_{\rm{CW}} = ( K-3J)/4$
(for $S=1/2$).  Since we obtained an estimation of the saturation
field and the ordering temperature from MC simulation for each
magnetic phase, we can estimate the actual values of $J$ and $K$ based
on these three parameters and speculate which ordered phase is
realized in the real material.  Table.\ref{tab:1} shows the ratio of
Curie-Weiss temperature $\Theta_{\rm{CW}}$ to angle-averaged
saturation field $h_{\rm{sat}}^{\rm{ave}}$ for different ordered
states (details are discussed in Sec.\ref{sec:magnetic-field}).  From
the observation of the saturation fields listed in Table.\ref{tab:1},
one can point out that in the N\'eel and skew-zig-zag phase, the
system can have small $h_{\rm{sat}}^{\rm{ave}}$ even when exchange
interactions $J$ and $K$ are large, due to the cancellation between
them.  On the other hand, if the system is in the skew-stripy phase,
the saturation field can serve as an estimation of $J$.

In summary, we studied the finite temperature phase diagram of the
Heisenberg-Kitaev (HK) model on a 3D hyperhoneycomb lattice. Based on
classical MC simulation and a semi-classical analysis using
Holstein-Primakoff bosons, we investigated the magnetic ordering
temperatures and ObD effects in the HK model. Unlike the case for the
2D HK model on the honeycomb lattice, the ordering temperature is
finite even in the limit of the pure (anti-) ferromagnetic Heisenberg
points $K/|J| = 0$ and their equivalent points $K/|J| = \pm 2$ for
$\pm J >0$ after the four-sublattice rotation.  In addition, the
overall energy scale of ordering temperatures in the HK model on the
3D hyperhoneycomb lattice is higher than the ones on the 2D honeycomb
lattice as expected.\cite{PhysRevB.88.024410} Based on the MC
simulation, one can see that the ordering temperature is largely
suppressed at the transition point $K/|J| = \pm1$ for $\pm J >0$ due
to the magnetic frustration.  Away from the transition point, however,
the ordering temperature $T/|J|$ increases with increase of $K/|J|$.
Linear spin-wave theory also shows the same trend of ordering
temperature and it reasonably estimates the ordering temperature
compare to the MC simulation.  We also studied ObD effects from both
thermal and quantum fluctuations. These two different types of
fluctuations compete with each other and they favor different magnetic
orders in a certain parameter region.  Just below the ordering
temperature, the entropy effect is large and the states selected by
thermal ObD are favored.  At very low temperatures, however,
zero-point quantum fluctuations are dominant and the states selected
by quantum ObD are favored.  Such competition between thermal ObD and
quantum ObD results in an additional phase transition below the
ordering temperature.  Finally, we investigated the magnetization
process and the saturation field as a function of $J$ and $K$ for each
phase.  For general field direction, we found that the spin-flop
transition is present in both skew-stripy phase and skew-zig-zag
phase. Such spin-flop transitions originate from the non-coplanar
character of the general skew-stripy/skew-zig-zag spin order. For
comparison with future experiments, we have also investigated in
detail the directional dependence of the saturation field and
calculated the angle-averaged saturation field at zero-temperature.

\acknowledgements We thank S. Bhattacharjee and R. Schaffer for discussions. YBK and
AP wish to acknowledge the hospitality of MPIPKS, Dresden. This
research was supported by the NSERC, CIFAR, and Centre for Quantum
Materials at the University of Toronto.

\bibliography{ObD-hyper-honeycomb}

\begin{thebibliography}{29}%
\makeatletter
\providecommand \@ifxundefined [1]{%
 \@ifx{#1\undefined}
}%
\providecommand \@ifnum [1]{%
 \ifnum #1\expandafter \@firstoftwo
 \else \expandafter \@secondoftwo
 \fi
}%
\providecommand \@ifx [1]{%
 \ifx #1\expandafter \@firstoftwo
 \else \expandafter \@secondoftwo
 \fi
}%
\providecommand \natexlab [1]{#1}%
\providecommand \enquote  [1]{``#1''}%
\providecommand \bibnamefont  [1]{#1}%
\providecommand \bibfnamefont [1]{#1}%
\providecommand \citenamefont [1]{#1}%
\providecommand \href@noop [0]{\@secondoftwo}%
\providecommand \href [0]{\begingroup \@sanitize@url \@href}%
\providecommand \@href[1]{\@@startlink{#1}\@@href}%
\providecommand \@@href[1]{\endgroup#1\@@endlink}%
\providecommand \@sanitize@url [0]{\catcode `\\12\catcode `\$12\catcode
  `\&12\catcode `\#12\catcode `\^12\catcode `\_12\catcode `\%12\relax}%
\providecommand \@@startlink[1]{}%
\providecommand \@@endlink[0]{}%
\providecommand \url  [0]{\begingroup\@sanitize@url \@url }%
\providecommand \@url [1]{\endgroup\@href {#1}{\urlprefix }}%
\providecommand \urlprefix  [0]{URL }%
\providecommand \Eprint [0]{\href }%
\providecommand \doibase [0]{http://dx.doi.org/}%
\providecommand \selectlanguage [0]{\@gobble}%
\providecommand \bibinfo  [0]{\@secondoftwo}%
\providecommand \bibfield  [0]{\@secondoftwo}%
\providecommand \translation [1]{[#1]}%
\providecommand \BibitemOpen [0]{}%
\providecommand \bibitemStop [0]{}%
\providecommand \bibitemNoStop [0]{.\EOS\space}%
\providecommand \EOS [0]{\spacefactor3000\relax}%
\providecommand \BibitemShut  [1]{\csname bibitem#1\endcsname}%
\let\auto@bib@innerbib\@empty
\bibitem [{\citenamefont {Witczak-Krempa}\ \emph {et~al.}(2013)\citenamefont
  {Witczak-Krempa}, \citenamefont {Chen}, \citenamefont {Kim},\ and\
  \citenamefont {Balents}}]{witczak2013correlated}%
  \BibitemOpen
  \bibfield  {author} {\bibinfo {author} {\bibfnamefont {W.}~\bibnamefont
  {Witczak-Krempa}}, \bibinfo {author} {\bibfnamefont {G.}~\bibnamefont
  {Chen}}, \bibinfo {author} {\bibfnamefont {Y.~B.}\ \bibnamefont {Kim}}, \
  and\ \bibinfo {author} {\bibfnamefont {L.}~\bibnamefont {Balents}},\
  }\href@noop {} {\bibfield  {journal} {\bibinfo  {journal} {arXiv preprint
  arXiv:1305.2193}\ } (\bibinfo {year} {2013})}\BibitemShut {NoStop}%
\bibitem [{\citenamefont {Singh}\ and\ \citenamefont
  {Gegenwart}(2010)}]{PhysRevB.82.064412}%
  \BibitemOpen
  \bibfield  {author} {\bibinfo {author} {\bibfnamefont {Y.}~\bibnamefont
  {Singh}}\ and\ \bibinfo {author} {\bibfnamefont {P.}~\bibnamefont
  {Gegenwart}},\ }\href {\doibase 10.1103/PhysRevB.82.064412} {\bibfield
  {journal} {\bibinfo  {journal} {Phys. Rev. B}\ }\textbf {\bibinfo {volume}
  {82}},\ \bibinfo {pages} {064412} (\bibinfo {year} {2010})}\BibitemShut
  {NoStop}%
\bibitem [{\citenamefont {Singh}\ \emph {et~al.}(2012)\citenamefont {Singh},
  \citenamefont {Manni}, \citenamefont {Reuther}, \citenamefont {Berlijn},
  \citenamefont {Thomale}, \citenamefont {Ku}, \citenamefont {Trebst},\ and\
  \citenamefont {Gegenwart}}]{PhysRevLett.108.127203}%
  \BibitemOpen
  \bibfield  {author} {\bibinfo {author} {\bibfnamefont {Y.}~\bibnamefont
  {Singh}}, \bibinfo {author} {\bibfnamefont {S.}~\bibnamefont {Manni}},
  \bibinfo {author} {\bibfnamefont {J.}~\bibnamefont {Reuther}}, \bibinfo
  {author} {\bibfnamefont {T.}~\bibnamefont {Berlijn}}, \bibinfo {author}
  {\bibfnamefont {R.}~\bibnamefont {Thomale}}, \bibinfo {author} {\bibfnamefont
  {W.}~\bibnamefont {Ku}}, \bibinfo {author} {\bibfnamefont {S.}~\bibnamefont
  {Trebst}}, \ and\ \bibinfo {author} {\bibfnamefont {P.}~\bibnamefont
  {Gegenwart}},\ }\href {\doibase 10.1103/PhysRevLett.108.127203} {\bibfield
  {journal} {\bibinfo  {journal} {Phys. Rev. Lett.}\ }\textbf {\bibinfo
  {volume} {108}},\ \bibinfo {pages} {127203} (\bibinfo {year}
  {2012})}\BibitemShut {NoStop}%
\bibitem [{\citenamefont {Gretarsson}\ \emph {et~al.}(2013)\citenamefont
  {Gretarsson}, \citenamefont {Clancy}, \citenamefont {Liu}, \citenamefont
  {Hill}, \citenamefont {Bozin}, \citenamefont {Singh}, \citenamefont {Manni},
  \citenamefont {Gegenwart}, \citenamefont {Kim}, \citenamefont {Said},
  \citenamefont {Casa}, \citenamefont {Gog}, \citenamefont {Upton},
  \citenamefont {Kim}, \citenamefont {Yu}, \citenamefont {Katukuri},
  \citenamefont {Hozoi}, \citenamefont {van~den Brink},\ and\ \citenamefont
  {Kim}}]{PhysRevLett.110.076402}%
  \BibitemOpen
  \bibfield  {author} {\bibinfo {author} {\bibfnamefont {H.}~\bibnamefont
  {Gretarsson}}, \bibinfo {author} {\bibfnamefont {J.~P.}\ \bibnamefont
  {Clancy}}, \bibinfo {author} {\bibfnamefont {X.}~\bibnamefont {Liu}},
  \bibinfo {author} {\bibfnamefont {J.~P.}\ \bibnamefont {Hill}}, \bibinfo
  {author} {\bibfnamefont {E.}~\bibnamefont {Bozin}}, \bibinfo {author}
  {\bibfnamefont {Y.}~\bibnamefont {Singh}}, \bibinfo {author} {\bibfnamefont
  {S.}~\bibnamefont {Manni}}, \bibinfo {author} {\bibfnamefont
  {P.}~\bibnamefont {Gegenwart}}, \bibinfo {author} {\bibfnamefont
  {J.}~\bibnamefont {Kim}}, \bibinfo {author} {\bibfnamefont {A.~H.}\
  \bibnamefont {Said}}, \bibinfo {author} {\bibfnamefont {D.}~\bibnamefont
  {Casa}}, \bibinfo {author} {\bibfnamefont {T.}~\bibnamefont {Gog}}, \bibinfo
  {author} {\bibfnamefont {M.~H.}\ \bibnamefont {Upton}}, \bibinfo {author}
  {\bibfnamefont {H.-S.}\ \bibnamefont {Kim}}, \bibinfo {author} {\bibfnamefont
  {J.}~\bibnamefont {Yu}}, \bibinfo {author} {\bibfnamefont {V.~M.}\
  \bibnamefont {Katukuri}}, \bibinfo {author} {\bibfnamefont {L.}~\bibnamefont
  {Hozoi}}, \bibinfo {author} {\bibfnamefont {J.}~\bibnamefont {van~den
  Brink}}, \ and\ \bibinfo {author} {\bibfnamefont {Y.-J.}\ \bibnamefont
  {Kim}},\ }\href {\doibase 10.1103/PhysRevLett.110.076402} {\bibfield
  {journal} {\bibinfo  {journal} {Phys. Rev. Lett.}\ }\textbf {\bibinfo
  {volume} {110}},\ \bibinfo {pages} {076402} (\bibinfo {year}
  {2013})}\BibitemShut {NoStop}%
\bibitem [{\citenamefont {Liu}\ \emph {et~al.}(2011)\citenamefont {Liu},
  \citenamefont {Berlijn}, \citenamefont {Yin}, \citenamefont {Ku},
  \citenamefont {Tsvelik}, \citenamefont {Kim}, \citenamefont {Gretarsson},
  \citenamefont {Singh}, \citenamefont {Gegenwart},\ and\ \citenamefont
  {Hill}}]{PhysRevB.83.220403}%
  \BibitemOpen
  \bibfield  {author} {\bibinfo {author} {\bibfnamefont {X.}~\bibnamefont
  {Liu}}, \bibinfo {author} {\bibfnamefont {T.}~\bibnamefont {Berlijn}},
  \bibinfo {author} {\bibfnamefont {W.-G.}\ \bibnamefont {Yin}}, \bibinfo
  {author} {\bibfnamefont {W.}~\bibnamefont {Ku}}, \bibinfo {author}
  {\bibfnamefont {A.}~\bibnamefont {Tsvelik}}, \bibinfo {author} {\bibfnamefont
  {Y.-J.}\ \bibnamefont {Kim}}, \bibinfo {author} {\bibfnamefont
  {H.}~\bibnamefont {Gretarsson}}, \bibinfo {author} {\bibfnamefont
  {Y.}~\bibnamefont {Singh}}, \bibinfo {author} {\bibfnamefont
  {P.}~\bibnamefont {Gegenwart}}, \ and\ \bibinfo {author} {\bibfnamefont
  {J.~P.}\ \bibnamefont {Hill}},\ }\href {\doibase 10.1103/PhysRevB.83.220403}
  {\bibfield  {journal} {\bibinfo  {journal} {Phys. Rev. B}\ }\textbf {\bibinfo
  {volume} {83}},\ \bibinfo {pages} {220403} (\bibinfo {year}
  {2011})}\BibitemShut {NoStop}%
\bibitem [{\citenamefont {Clancy}\ \emph {et~al.}(2012)\citenamefont {Clancy},
  \citenamefont {Chen}, \citenamefont {Kim}, \citenamefont {Chen},
  \citenamefont {Plumb}, \citenamefont {Jeon}, \citenamefont {Noh},\ and\
  \citenamefont {Kim}}]{PhysRevB.86.195131}%
  \BibitemOpen
  \bibfield  {author} {\bibinfo {author} {\bibfnamefont {J.~P.}\ \bibnamefont
  {Clancy}}, \bibinfo {author} {\bibfnamefont {N.}~\bibnamefont {Chen}},
  \bibinfo {author} {\bibfnamefont {C.~Y.}\ \bibnamefont {Kim}}, \bibinfo
  {author} {\bibfnamefont {W.~F.}\ \bibnamefont {Chen}}, \bibinfo {author}
  {\bibfnamefont {K.~W.}\ \bibnamefont {Plumb}}, \bibinfo {author}
  {\bibfnamefont {B.~C.}\ \bibnamefont {Jeon}}, \bibinfo {author}
  {\bibfnamefont {T.~W.}\ \bibnamefont {Noh}}, \ and\ \bibinfo {author}
  {\bibfnamefont {Y.-J.}\ \bibnamefont {Kim}},\ }\href {\doibase
  10.1103/PhysRevB.86.195131} {\bibfield  {journal} {\bibinfo  {journal} {Phys.
  Rev. B}\ }\textbf {\bibinfo {volume} {86}},\ \bibinfo {pages} {195131}
  (\bibinfo {year} {2012})}\BibitemShut {NoStop}%
\bibitem [{\citenamefont {Choi}\ \emph {et~al.}(2012)\citenamefont {Choi},
  \citenamefont {Coldea}, \citenamefont {Kolmogorov}, \citenamefont
  {Lancaster}, \citenamefont {Mazin}, \citenamefont {Blundell}, \citenamefont
  {Radaelli}, \citenamefont {Singh}, \citenamefont {Gegenwart}, \citenamefont
  {Choi}, \citenamefont {Cheong}, \citenamefont {Baker}, \citenamefont
  {Stock},\ and\ \citenamefont {Taylor}}]{PhysRevLett.108.127204}%
  \BibitemOpen
  \bibfield  {author} {\bibinfo {author} {\bibfnamefont {S.~K.}\ \bibnamefont
  {Choi}}, \bibinfo {author} {\bibfnamefont {R.}~\bibnamefont {Coldea}},
  \bibinfo {author} {\bibfnamefont {A.~N.}\ \bibnamefont {Kolmogorov}},
  \bibinfo {author} {\bibfnamefont {T.}~\bibnamefont {Lancaster}}, \bibinfo
  {author} {\bibfnamefont {I.~I.}\ \bibnamefont {Mazin}}, \bibinfo {author}
  {\bibfnamefont {S.~J.}\ \bibnamefont {Blundell}}, \bibinfo {author}
  {\bibfnamefont {P.~G.}\ \bibnamefont {Radaelli}}, \bibinfo {author}
  {\bibfnamefont {Y.}~\bibnamefont {Singh}}, \bibinfo {author} {\bibfnamefont
  {P.}~\bibnamefont {Gegenwart}}, \bibinfo {author} {\bibfnamefont {K.~R.}\
  \bibnamefont {Choi}}, \bibinfo {author} {\bibfnamefont {S.-W.}\ \bibnamefont
  {Cheong}}, \bibinfo {author} {\bibfnamefont {P.~J.}\ \bibnamefont {Baker}},
  \bibinfo {author} {\bibfnamefont {C.}~\bibnamefont {Stock}}, \ and\ \bibinfo
  {author} {\bibfnamefont {J.}~\bibnamefont {Taylor}},\ }\href {\doibase
  10.1103/PhysRevLett.108.127204} {\bibfield  {journal} {\bibinfo  {journal}
  {Phys. Rev. Lett.}\ }\textbf {\bibinfo {volume} {108}},\ \bibinfo {pages}
  {127204} (\bibinfo {year} {2012})}\BibitemShut {NoStop}%
\bibitem [{\citenamefont {Jackeli}\ and\ \citenamefont
  {Khaliullin}(2009)}]{PhysRevLett.102.017205}%
  \BibitemOpen
  \bibfield  {author} {\bibinfo {author} {\bibfnamefont {G.}~\bibnamefont
  {Jackeli}}\ and\ \bibinfo {author} {\bibfnamefont {G.}~\bibnamefont
  {Khaliullin}},\ }\href {\doibase 10.1103/PhysRevLett.102.017205} {\bibfield
  {journal} {\bibinfo  {journal} {Phys. Rev. Lett.}\ }\textbf {\bibinfo
  {volume} {102}},\ \bibinfo {pages} {017205} (\bibinfo {year}
  {2009})}\BibitemShut {NoStop}%
\bibitem [{\citenamefont {Kitaev}(2006)}]{Kitaev20062}%
  \BibitemOpen
  \bibfield  {author} {\bibinfo {author} {\bibfnamefont {A.}~\bibnamefont
  {Kitaev}},\ }\href {\doibase http://dx.doi.org/10.1016/j.aop.2005.10.005}
  {\bibfield  {journal} {\bibinfo  {journal} {Annals of Physics}\ }\textbf
  {\bibinfo {volume} {321}},\ \bibinfo {pages} {2 } (\bibinfo {year}
  {2006})}\BibitemShut {NoStop}%
\bibitem [{\citenamefont {Mandal}\ and\ \citenamefont
  {Surendran}(2009)}]{PhysRevB.79.024426}%
  \BibitemOpen
  \bibfield  {author} {\bibinfo {author} {\bibfnamefont {S.}~\bibnamefont
  {Mandal}}\ and\ \bibinfo {author} {\bibfnamefont {N.}~\bibnamefont
  {Surendran}},\ }\href {\doibase 10.1103/PhysRevB.79.024426} {\bibfield
  {journal} {\bibinfo  {journal} {Phys. Rev. B}\ }\textbf {\bibinfo {volume}
  {79}},\ \bibinfo {pages} {024426} (\bibinfo {year} {2009})}\BibitemShut
  {NoStop}%
\bibitem [{\citenamefont {Chaloupka}\ \emph {et~al.}(2010)\citenamefont
  {Chaloupka}, \citenamefont {Jackeli},\ and\ \citenamefont
  {Khaliullin}}]{PhysRevLett.105.027204}%
  \BibitemOpen
  \bibfield  {author} {\bibinfo {author} {\bibfnamefont {J.~c.~v.}\
  \bibnamefont {Chaloupka}}, \bibinfo {author} {\bibfnamefont {G.}~\bibnamefont
  {Jackeli}}, \ and\ \bibinfo {author} {\bibfnamefont {G.}~\bibnamefont
  {Khaliullin}},\ }\href {\doibase 10.1103/PhysRevLett.105.027204} {\bibfield
  {journal} {\bibinfo  {journal} {Phys. Rev. Lett.}\ }\textbf {\bibinfo
  {volume} {105}},\ \bibinfo {pages} {027204} (\bibinfo {year}
  {2010})}\BibitemShut {NoStop}%
\bibitem [{\citenamefont {Kimchi}\ and\ \citenamefont
  {You}(2011)}]{PhysRevB.84.180407}%
  \BibitemOpen
  \bibfield  {author} {\bibinfo {author} {\bibfnamefont {I.}~\bibnamefont
  {Kimchi}}\ and\ \bibinfo {author} {\bibfnamefont {Y.-Z.}\ \bibnamefont
  {You}},\ }\href {\doibase 10.1103/PhysRevB.84.180407} {\bibfield  {journal}
  {\bibinfo  {journal} {Phys. Rev. B}\ }\textbf {\bibinfo {volume} {84}},\
  \bibinfo {pages} {180407} (\bibinfo {year} {2011})}\BibitemShut {NoStop}%
\bibitem [{\citenamefont {Mazin}\ \emph {et~al.}(2012)\citenamefont {Mazin},
  \citenamefont {Jeschke}, \citenamefont {Foyevtsova}, \citenamefont
  {Valent\'\i},\ and\ \citenamefont {Khomskii}}]{PhysRevLett.109.197201}%
  \BibitemOpen
  \bibfield  {author} {\bibinfo {author} {\bibfnamefont {I.~I.}\ \bibnamefont
  {Mazin}}, \bibinfo {author} {\bibfnamefont {H.~O.}\ \bibnamefont {Jeschke}},
  \bibinfo {author} {\bibfnamefont {K.}~\bibnamefont {Foyevtsova}}, \bibinfo
  {author} {\bibfnamefont {R.}~\bibnamefont {Valent\'\i}}, \ and\ \bibinfo
  {author} {\bibfnamefont {D.~I.}\ \bibnamefont {Khomskii}},\ }\href {\doibase
  10.1103/PhysRevLett.109.197201} {\bibfield  {journal} {\bibinfo  {journal}
  {Phys. Rev. Lett.}\ }\textbf {\bibinfo {volume} {109}},\ \bibinfo {pages}
  {197201} (\bibinfo {year} {2012})}\BibitemShut {NoStop}%
\bibitem [{\citenamefont {Foyevtsova}\ \emph {et~al.}(2013)\citenamefont
  {Foyevtsova}, \citenamefont {Jeschke}, \citenamefont {Mazin}, \citenamefont
  {Khomskii},\ and\ \citenamefont {Valent\'\i}}]{PhysRevB.88.035107}%
  \BibitemOpen
  \bibfield  {author} {\bibinfo {author} {\bibfnamefont {K.}~\bibnamefont
  {Foyevtsova}}, \bibinfo {author} {\bibfnamefont {H.~O.}\ \bibnamefont
  {Jeschke}}, \bibinfo {author} {\bibfnamefont {I.~I.}\ \bibnamefont {Mazin}},
  \bibinfo {author} {\bibfnamefont {D.~I.}\ \bibnamefont {Khomskii}}, \ and\
  \bibinfo {author} {\bibfnamefont {R.}~\bibnamefont {Valent\'\i}},\ }\href
  {\doibase 10.1103/PhysRevB.88.035107} {\bibfield  {journal} {\bibinfo
  {journal} {Phys. Rev. B}\ }\textbf {\bibinfo {volume} {88}},\ \bibinfo
  {pages} {035107} (\bibinfo {year} {2013})}\BibitemShut {NoStop}%
\bibitem [{\citenamefont {Takagi}(2013)}]{2013_takagi}%
  \BibitemOpen
  \bibfield  {author} {\bibinfo {author} {\bibfnamefont {H.}~\bibnamefont
  {Takagi}},\ }\href@noop {} {\bibfield  {journal} {\bibinfo  {journal} {Talk
  in Conference on {\it Spin Orbit Entanglement: Exotic States of Quantum
  Matter in Electronic Systems}, MPIPKS, Dresden,}\ } (\bibinfo {year} {July
  22-July 26, 2013})}\BibitemShut {NoStop}%
\bibitem [{\citenamefont {Lee}\ \emph {et~al.}(2013)\citenamefont {Lee},
  \citenamefont {Schaffer}, \citenamefont {Bhattacharjee},\ and\ \citenamefont
  {Kim}}]{lee2013heisenberg}%
  \BibitemOpen
  \bibfield  {author} {\bibinfo {author} {\bibfnamefont {E.~K.-H.}\
  \bibnamefont {Lee}}, \bibinfo {author} {\bibfnamefont {R.}~\bibnamefont
  {Schaffer}}, \bibinfo {author} {\bibfnamefont {S.}~\bibnamefont
  {Bhattacharjee}}, \ and\ \bibinfo {author} {\bibfnamefont {Y.~B.}\
  \bibnamefont {Kim}},\ }\href@noop {} {\bibfield  {journal} {\bibinfo
  {journal} {arXiv preprint arXiv:1308.6592}\ } (\bibinfo {year}
  {2013})}\BibitemShut {NoStop}%
\bibitem [{\citenamefont {Kimchi}\ \emph {et~al.}(2013)\citenamefont {Kimchi},
  \citenamefont {Analytis},\ and\ \citenamefont
  {Vishwanath}}]{kimchi2013three}%
  \BibitemOpen
  \bibfield  {author} {\bibinfo {author} {\bibfnamefont {I.}~\bibnamefont
  {Kimchi}}, \bibinfo {author} {\bibfnamefont {J.~G.}\ \bibnamefont
  {Analytis}}, \ and\ \bibinfo {author} {\bibfnamefont {A.}~\bibnamefont
  {Vishwanath}},\ }\href@noop {} {\bibfield  {journal} {\bibinfo  {journal}
  {arXiv preprint arXiv:1309.1171}\ } (\bibinfo {year} {2013})}\BibitemShut
  {NoStop}%
\bibitem [{\citenamefont {Nasu}\ \emph {et~al.}(2013)\citenamefont {Nasu},
  \citenamefont {Kaji}, \citenamefont {Matsuura}, \citenamefont {Udagawa},\
  and\ \citenamefont {Motome}}]{nasu2013finite}%
  \BibitemOpen
  \bibfield  {author} {\bibinfo {author} {\bibfnamefont {J.}~\bibnamefont
  {Nasu}}, \bibinfo {author} {\bibfnamefont {T.}~\bibnamefont {Kaji}}, \bibinfo
  {author} {\bibfnamefont {K.}~\bibnamefont {Matsuura}}, \bibinfo {author}
  {\bibfnamefont {M.}~\bibnamefont {Udagawa}}, \ and\ \bibinfo {author}
  {\bibfnamefont {Y.}~\bibnamefont {Motome}},\ }\href@noop {} {\bibfield
  {journal} {\bibinfo  {journal} {arXiv preprint arXiv:1309.3068}\ } (\bibinfo
  {year} {2013})}\BibitemShut {NoStop}%
\bibitem [{\citenamefont {Price}\ and\ \citenamefont
  {Perkins}(2013)}]{PhysRevB.88.024410}%
  \BibitemOpen
  \bibfield  {author} {\bibinfo {author} {\bibfnamefont {C.}~\bibnamefont
  {Price}}\ and\ \bibinfo {author} {\bibfnamefont {N.~B.}\ \bibnamefont
  {Perkins}},\ }\href {\doibase 10.1103/PhysRevB.88.024410} {\bibfield
  {journal} {\bibinfo  {journal} {Phys. Rev. B}\ }\textbf {\bibinfo {volume}
  {88}},\ \bibinfo {pages} {024410} (\bibinfo {year} {2013})}\BibitemShut
  {NoStop}%
\bibitem [{\citenamefont {Bergman}\ \emph {et~al.}(2007)\citenamefont
  {Bergman}, \citenamefont {Alicea}, \citenamefont {Gull}, \citenamefont
  {Trebst},\ and\ \citenamefont {Balents}}]{bergman2007order}%
  \BibitemOpen
  \bibfield  {author} {\bibinfo {author} {\bibfnamefont {D.}~\bibnamefont
  {Bergman}}, \bibinfo {author} {\bibfnamefont {J.}~\bibnamefont {Alicea}},
  \bibinfo {author} {\bibfnamefont {E.}~\bibnamefont {Gull}}, \bibinfo {author}
  {\bibfnamefont {S.}~\bibnamefont {Trebst}}, \ and\ \bibinfo {author}
  {\bibfnamefont {L.}~\bibnamefont {Balents}},\ }\href@noop {} {\bibfield
  {journal} {\bibinfo  {journal} {Nature Physics}\ }\textbf {\bibinfo {volume}
  {3}},\ \bibinfo {pages} {487} (\bibinfo {year} {2007})}\BibitemShut {NoStop}%
\bibitem [{\citenamefont {Bernier}\ \emph {et~al.}(2008)\citenamefont
  {Bernier}, \citenamefont {Lawler},\ and\ \citenamefont
  {Kim}}]{bernier2008quantum}%
  \BibitemOpen
  \bibfield  {author} {\bibinfo {author} {\bibfnamefont {J.-S.}\ \bibnamefont
  {Bernier}}, \bibinfo {author} {\bibfnamefont {M.~J.}\ \bibnamefont {Lawler}},
  \ and\ \bibinfo {author} {\bibfnamefont {Y.~B.}\ \bibnamefont {Kim}},\ }\href
  {\doibase 10.1103/PhysRevLett.101.047201} {\bibfield  {journal} {\bibinfo
  {journal} {Phys. Rev. Lett.}\ }\textbf {\bibinfo {volume} {101}},\ \bibinfo
  {pages} {047201} (\bibinfo {year} {2008})}\BibitemShut {NoStop}%
\bibitem [{\citenamefont {Chern}(2010)}]{chern2010pyrochlore}%
  \BibitemOpen
  \bibfield  {author} {\bibinfo {author} {\bibfnamefont {G.-W.}\ \bibnamefont
  {Chern}},\ }\href@noop {} {\bibfield  {journal} {\bibinfo  {journal} {arXiv
  preprint arXiv:1008.3038}\ } (\bibinfo {year} {2010})}\BibitemShut {NoStop}%
\bibitem [{\citenamefont {Kim}\ \emph {et~al.}(2009)\citenamefont {Kim},
  \citenamefont {Ohsumi}, \citenamefont {Komesu}, \citenamefont {Sakai},
  \citenamefont {Morita}, \citenamefont {Takagi},\ and\ \citenamefont
  {Arima}}]{kim06032009}%
  \BibitemOpen
  \bibfield  {author} {\bibinfo {author} {\bibfnamefont {B.~J.}\ \bibnamefont
  {Kim}}, \bibinfo {author} {\bibfnamefont {H.}~\bibnamefont {Ohsumi}},
  \bibinfo {author} {\bibfnamefont {T.}~\bibnamefont {Komesu}}, \bibinfo
  {author} {\bibfnamefont {S.}~\bibnamefont {Sakai}}, \bibinfo {author}
  {\bibfnamefont {T.}~\bibnamefont {Morita}}, \bibinfo {author} {\bibfnamefont
  {H.}~\bibnamefont {Takagi}}, \ and\ \bibinfo {author} {\bibfnamefont
  {T.}~\bibnamefont {Arima}},\ }\href {\doibase 10.1126/science.1167106}
  {\bibfield  {journal} {\bibinfo  {journal} {Science}\ }\textbf {\bibinfo
  {volume} {323}},\ \bibinfo {pages} {1329} (\bibinfo {year}
  {2009})}\BibitemShut {NoStop}%
\bibitem [{\citenamefont {Witczak-Krempa}\ and\ \citenamefont
  {Kim}(2012)}]{witczak2012topological}%
  \BibitemOpen
  \bibfield  {author} {\bibinfo {author} {\bibfnamefont {W.}~\bibnamefont
  {Witczak-Krempa}}\ and\ \bibinfo {author} {\bibfnamefont {Y.~B.}\
  \bibnamefont {Kim}},\ }\href {\doibase 10.1103/PhysRevB.85.045124} {\bibfield
   {journal} {\bibinfo  {journal} {Phys. Rev. B}\ }\textbf {\bibinfo {volume}
  {85}},\ \bibinfo {pages} {045124} (\bibinfo {year} {2012})}\BibitemShut
  {NoStop}%
\bibitem [{\citenamefont {Khaliullin}(2005)}]{khaliullin2005orbital}%
  \BibitemOpen
  \bibfield  {author} {\bibinfo {author} {\bibfnamefont {G.}~\bibnamefont
  {Khaliullin}},\ }\href@noop {} {\bibfield  {journal} {\bibinfo  {journal}
  {Progress of Theoretical Physics Supplement}\ }\textbf {\bibinfo {volume}
  {160}},\ \bibinfo {pages} {155} (\bibinfo {year} {2005})}\BibitemShut
  {NoStop}%
\bibitem [{\citenamefont {Caselle}\ and\ \citenamefont
  {Hasenbusch}(1998)}]{caselle1998stability}%
  \BibitemOpen
  \bibfield  {author} {\bibinfo {author} {\bibfnamefont {M.}~\bibnamefont
  {Caselle}}\ and\ \bibinfo {author} {\bibfnamefont {M.}~\bibnamefont
  {Hasenbusch}},\ }\href@noop {} {\bibfield  {journal} {\bibinfo  {journal}
  {Journal of Physics A: Mathematical and General}\ }\textbf {\bibinfo {volume}
  {31}},\ \bibinfo {pages} {4603} (\bibinfo {year} {1998})}\BibitemShut
  {NoStop}%
\bibitem [{\citenamefont {Scholten}\ and\ \citenamefont
  {Irakliotis}(1993)}]{PhysRevB.48.1291}%
  \BibitemOpen
  \bibfield  {author} {\bibinfo {author} {\bibfnamefont {P.~D.}\ \bibnamefont
  {Scholten}}\ and\ \bibinfo {author} {\bibfnamefont {L.~J.}\ \bibnamefont
  {Irakliotis}},\ }\href {\doibase 10.1103/PhysRevB.48.1291} {\bibfield
  {journal} {\bibinfo  {journal} {Phys. Rev. B}\ }\textbf {\bibinfo {volume}
  {48}},\ \bibinfo {pages} {1291} (\bibinfo {year} {1993})}\BibitemShut
  {NoStop}%
\bibitem [{\citenamefont {Lou}\ \emph {et~al.}(2007)\citenamefont {Lou},
  \citenamefont {Sandvik},\ and\ \citenamefont
  {Balents}}]{PhysRevLett.99.207203}%
  \BibitemOpen
  \bibfield  {author} {\bibinfo {author} {\bibfnamefont {J.}~\bibnamefont
  {Lou}}, \bibinfo {author} {\bibfnamefont {A.~W.}\ \bibnamefont {Sandvik}}, \
  and\ \bibinfo {author} {\bibfnamefont {L.}~\bibnamefont {Balents}},\ }\href
  {\doibase 10.1103/PhysRevLett.99.207203} {\bibfield  {journal} {\bibinfo
  {journal} {Phys. Rev. Lett.}\ }\textbf {\bibinfo {volume} {99}},\ \bibinfo
  {pages} {207203} (\bibinfo {year} {2007})}\BibitemShut {NoStop}%
\bibitem [{\citenamefont {Hove}\ and\ \citenamefont
  {Sudb\o{}}(2003)}]{PhysRevE.68.046107}%
  \BibitemOpen
  \bibfield  {author} {\bibinfo {author} {\bibfnamefont {J.}~\bibnamefont
  {Hove}}\ and\ \bibinfo {author} {\bibfnamefont {A.}~\bibnamefont
  {Sudb\o{}}},\ }\href {\doibase 10.1103/PhysRevE.68.046107} {\bibfield
  {journal} {\bibinfo  {journal} {Phys. Rev. E}\ }\textbf {\bibinfo {volume}
  {68}},\ \bibinfo {pages} {046107} (\bibinfo {year} {2003})}\BibitemShut
  {NoStop}%
\end{thebibliography}%

\newpage

\section{Appendix}
\label{sec:appendix}

\subsection{Free energy calculation}
\label{app:free-energy}

In this Appendix, we discuss the free energy calculation in the ferromagnetic phase for $K/|J| >1$ and $J<0$ 
and find the state selected by thermal ObD.
Let us start from a ferromagnetic ground state and expand fluctuations by writing
\bea
\bS_i &=& \boldsymbol{\bar{S}} \sqrt{1- \Pi_i^2} + \boldsymbol{\Pi}_i ~, \\
\boldsymbol{\Pi}_i &=& \boldsymbol{v}_1 \phi_i + \boldsymbol{v}_2 \chi_i ~,
\eea
where $\boldsymbol{\bar{S}}$ is FM order pointing along certain direction 
and $\boldsymbol{\Pi}_i$ is lying on its basal plane satisfying 
$\boldsymbol{\bar{S}} \cdot \boldsymbol{\Pi}_i =0$.  
This automatically satisfies a constraint $| \bS_i | =1$ and the partition function becomes 
\begin{widetext}
\bea
\mathcal{Z} &=&  \int \mathcal{D} \bS ~ e^{-\beta H }
= \int \mathcal{D} \chi \mathcal{D} \phi ~e^{-\mathcal{S}}  ~,  \\
\mathcal{S} &=&  \frac{1}{2 T} 
\sum_{ij}~
 (\phi_i \tilde{J}_{ij} \phi_j  
+ \chi_i \tilde{J}_{ij} \chi_j   + \phi_i {P}_{ij}  \phi_j
+ \chi_i {Q}_{ij}   \chi_j 
+ 2 \phi_i R_{ij}  \chi_j ) \nonumber \\
&=& \frac{1}{2 T}
\sum_{\boldsymbol{k}}~
\left( \begin{array}{cc}
\phi_{\boldsymbol{k}}^* &
\chi_{\boldsymbol{k}}^*
\end{array} \right)
M_{\boldsymbol{k}}
\left( \begin{array}{cc}
\phi_{\boldsymbol{k}} \\
\chi_{\boldsymbol{k}} ~
\end{array} \right)~,
\label{eq:5}
\eea
\end{widetext}
where
\bea
\tilde{J}_{ij} &=& J_{ij} - 3J \delta_{ij} ,\nonumber \\ 
P_{ij} &=& -K_{ij}^\alpha( v_1^\alpha)^2+K \delta_{ij}, \nonumber \\
Q_{ij} &=& -K_{ij}^\alpha (v_2^\alpha)^2 +K \delta_{ij},  \nonumber\\ 
R_{ij} &=& -K_{ij}^\alpha (v_1^\alpha  v_2^\alpha) . 
\eea
Here, we keep only quadratic terms of $\chi $ and $\phi$ assuming that
this Gaussian action well behaves at low enough temperature.
The free energy $\mathcal{F}$ can be represented as  
\bea
F &=& - T~ {\rm{ln}}~(\mathcal{Z}) \nonumber \\
                     &=& \frac{T}{2} \sum_{\boldsymbol{k}, a} ~{\rm{ln}}~( W_a (\boldsymbol{k})/2\pi T ), 
\eea
where  $W_a (\boldsymbol{k})$ is the $a$-th eigenvalues of matrix $M_{\boldsymbol{k}}$
defined in Eq.\eqref{eq:5}, at a given wave vector $\boldsymbol{k}$.
The connection between this classical approach and 
a semi-classical analysis using Holstein-Primakoff bosons,
can be understood in the following way.
In high temperature limit, 
Eq.\eqref{eq:freeEnergyHp} (in the main text) can be rewritten as 
\bea
F(T) &=& T \sum_{ \boldsymbol{k} , \omega_{\boldsymbol{k}} >0}  {\rm{ln}} 
(1-e^{ - \omega_{\boldsymbol{k} }/T} )  \nonumber \\
         &\approx & T \sum_{\boldsymbol{k}, \omega_{\boldsymbol{k}} >0}
         {\rm{ln}}~\frac{\omega_{\boldsymbol{k}} }{T}.
\eea
The trace of ${\rm{ln}}~\omega_{\boldsymbol{k}} $ is consistent with 
trace of $1/2 ~ {\rm{ln}}~W_a (\boldsymbol{k})$ in Eq.\eqref{eq:5}
and in this way, the free energy of semi-classical analysis recovers 
the free energy of classical approach in high temperature limit.

\begin{figure}[t]
\scalebox{0.5}{\includegraphics{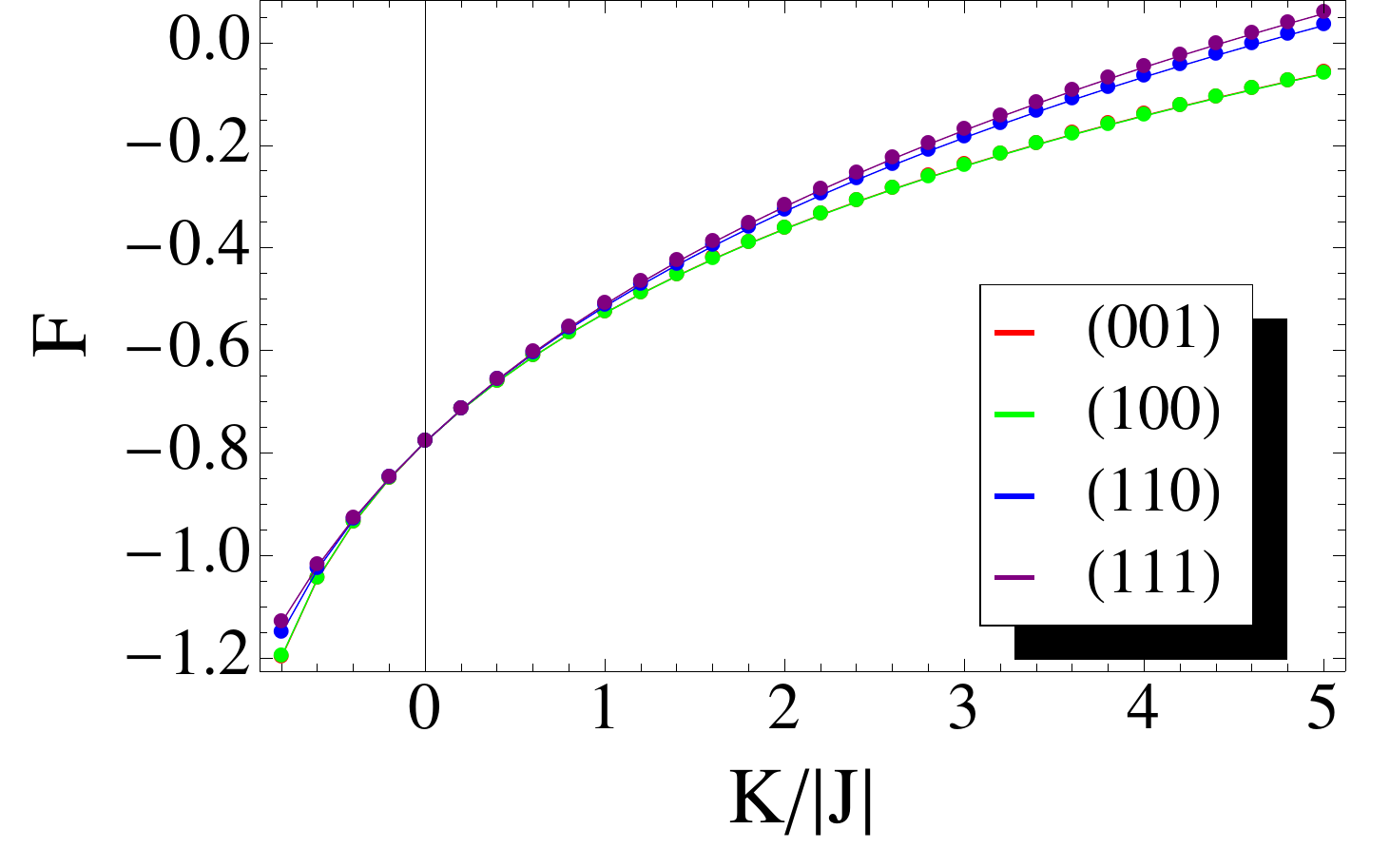}}
\caption{Calculated free energy from the quadratic order of fluctuating fields $\phi,\chi$
 for different FM ordering vectors : Red, green, blue and purple colored lines 
 are free energies for FM order 
 $\boldsymbol{\bar{S}}// (001),(100),(110),(111)$ respectively. Free energy differences between 
 $\boldsymbol{\bar{S}} = (001)$ and $\boldsymbol{\bar{S}} =(100)$ 
 are replotted in Fig.\ref{fig:app-2} for a better resolution.}
\label{fig:app-1}
\end{figure}

\begin{figure}[t]
\scalebox{0.6}{\includegraphics{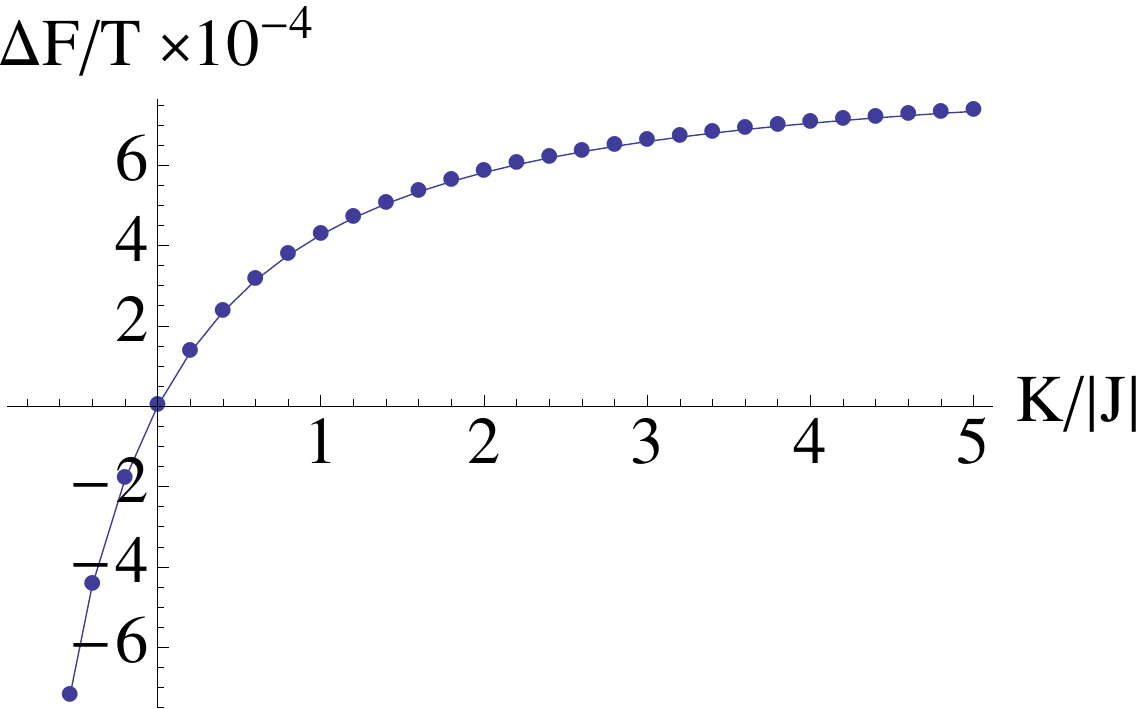}}
\caption{Plot of free energy difference between the two different FM order, 
$\Delta F /T = (F_{\boldsymbol{\bar{S}} = {(001)}} -F_{\boldsymbol{\bar{S}}= {(100)}})/T$.
Thermal ObD selects FM order along $(001)$ for $K/|J| < 0$, whereas,  
it favors FM order along $(100)$ for $K/|J| >0$. 
}
\label{fig:app-2}
\end{figure}

Fig.\ref{fig:app-1} shows the calculated free energy for different FM order with spins along 
$\ (001), (100),(110), (111)$ at $T/|J|=1$. 
The free energies of magnetic orders along $(001)$ or $(100)$ are always lower than the others  
for entire parameter regime of $K/|J|$ within FM order, resulting in either $(001)$- or $(100)$-FM states 
being selected by thermal fluctuations.  
The free energy difference between those two states are quite small, so we plot the energy difference 
$\Delta F /T = (F_{\boldsymbol{\bar{S}} = {(001)}} -F_{\boldsymbol{\bar{S}} = {(100)}})/T$ 
in Fig.\ref{fig:app-2}. 
As can be seen in Fig.\ref{fig:app-2}, the free energy difference $\Delta F/T$ changes its sign 
with the sign of Kitaev term $K$. 
For $K/|J| > 0$ where both Heisenberg and Kitaev interactions favor the FM order (unfrustrated case), 
thermal ObD selects the $(100)$-FM order.
On the other hand, for $K/|J| <0$ where Heisenberg and Kitaev interactions are frustrated, 
thermal ObD favors the $(001)$-FM order.
We also found that such thermal ObD effect exists in the N\'eel order.
For unfrustrated case (both $J$ and $K$ have same sign), thermal fluctuation favors
the N\'eel order with spins pointing along $(100)$, whereas, it favors the N\'eel order 
along $(001)$ for frustrated case (when $J$ and $K$ have different signs).
As emphasized in Sec.\ref{sec:HK},
the FM order for $J<0$ can be directly mapped onto the skew-stripy order for $J>0$, similarly 
the N\'eel order for $J>0$ onto the skew-zig-zag order for $J<0$, followed by a four-sublattices rotation
introduced in Sec.\ref{sec:HK}. 
Hence, one can expect the same thermal ObD effect for both skew-stripy phase and skew-zig-zag phase.

%
%

\end{document}